\pdfoutput=1
\documentclass[12pt,a4paper]{article}

\usepackage{ifthen} 
\usepackage{multirow}
\usepackage{pdflscape}
\usepackage{float}
\usepackage{booktabs}
\usepackage{rotating}
\usepackage{amsmath}
\usepackage{amssymb}
\usepackage{rotating}
\usepackage{verbatim}  
\usepackage[flushleft]{threeparttable}
\usepackage[usenames,dvipsnames]{xcolor}
\definecolor{taplum}{rgb}{0.67843, 0.49804, 0.65882}

\usepackage{bm} 
\usepackage{float} 
\usepackage{subfig} 

\newboolean{pdflatex}
\setboolean{pdflatex}{true} 

\newboolean{articletitles}
\setboolean{articletitles}{true} 

\newboolean{uprightparticles}
\setboolean{uprightparticles}{false} 

\newboolean{inbibliography}
\setboolean{inbibliography}{false} 

\usepackage[top=1in, bottom=1.25in, left=1in, right=1in]{geometry}

\usepackage{microtype}
\usepackage{lineno}  
\usepackage{xspace} 
\usepackage{caption} 

\usepackage{graphicx}  
\usepackage{color}
\usepackage{colortbl}
\graphicspath{{./figs/}} 

\usepackage{amsmath} 
\usepackage{amssymb}
\usepackage{amsfonts}
\usepackage{upgreek} 

\newcommand*\patchAmsMathEnvironmentForLineno[1]{%
\expandafter\let\csname old#1\expandafter\endcsname\csname #1\endcsname
\expandafter\let\csname oldend#1\expandafter\endcsname\csname
end#1\endcsname
 \renewenvironment{#1}%
   {\linenomath\csname old#1\endcsname}%
   {\csname oldend#1\endcsname\endlinenomath}%
}
\newcommand*\patchBothAmsMathEnvironmentsForLineno[1]{%
  \patchAmsMathEnvironmentForLineno{#1}%
  \patchAmsMathEnvironmentForLineno{#1*}%
}
\AtBeginDocument{%
\patchBothAmsMathEnvironmentsForLineno{equation}%
\patchBothAmsMathEnvironmentsForLineno{align}%
\patchBothAmsMathEnvironmentsForLineno{flalign}%
\patchBothAmsMathEnvironmentsForLineno{alignat}%
\patchBothAmsMathEnvironmentsForLineno{gather}%
\patchBothAmsMathEnvironmentsForLineno{multline}%
\patchBothAmsMathEnvironmentsForLineno{eqnarray}%
}

\usepackage{hyperref}    
\usepackage[all]{hypcap} 

\usepackage{xspace} 
\usepackage{upgreek}

\def\lhcb {\mbox{LHCb}\xspace}

\def\lhc    {\mbox{LHC}\xspace}

\def\MagUp {\mbox{\em Mag\kern -0.05em Up}\xspace}

\ifthenelse{\boolean{uprightparticles}}%
{

 \def\Ppi         {\ensuremath{\uppi}\xspace}

 \def\Ppsi        {\ensuremath{\uppsi}\xspace}

 \def\PDelta      {\ensuremath{\Delta}\xspace}                 
 \def\PXi      {\ensuremath{\Xi}\xspace}                 
 \def\PLambda      {\ensuremath{\Lambda}\xspace}                 
 \def\PSigma      {\ensuremath{\Sigma}\xspace}                 
 \def\POmega      {\ensuremath{\Omega}\xspace}                 
 \def\PUpsilon      {\ensuremath{\Upsilon}\xspace}                 
 
 \def\PB      {\ensuremath{\mathrm{B}}\xspace}                 
                  
 \def\PD      {\ensuremath{\mathrm{D}}\xspace}

 \def\PH      {\ensuremath{\mathrm{H}}\xspace}                 
                  
 \def\PJ      {\ensuremath{\mathrm{J}}\xspace}                 
 \def\PK      {\ensuremath{\mathrm{K}}\xspace}

 \def\Pb      {\ensuremath{\mathrm{b}}\xspace}                 
 \def\Pc      {\ensuremath{\mathrm{c}}\xspace}

 \def\Pi      {\ensuremath{\mathrm{i}}\xspace}

 \def\Pp      {\ensuremath{\mathrm{p}}\xspace}                 
 \def\Pq      {\ensuremath{\mathrm{q}}\xspace}

}
{

 \def\Ppi         {\ensuremath{\pi}\xspace}

 \def\Ppsi        {\ensuremath{\psi}\xspace}                 
                  
 \mathchardef\PDelta="7101
 \mathchardef\PXi="7104
 \mathchardef\PLambda="7103
 \mathchardef\PSigma="7106
 \mathchardef\POmega="710A
 \mathchardef\PUpsilon="7107
                  
 \def\PB      {\ensuremath{B}\xspace}                 
                  
 \def\PD      {\ensuremath{D}\xspace}

 \def\PH      {\ensuremath{H}\xspace}                 
                  
 \def\PJ      {\ensuremath{J}\xspace}                 
 \def\PK      {\ensuremath{K}\xspace}

 \def\Pb      {\ensuremath{b}\xspace}                 
 \def\Pc      {\ensuremath{c}\xspace}

 \def\Pi      {\ensuremath{i}\xspace}

 \def\Pp      {\ensuremath{p}\xspace}                 
 \def\Pq      {\ensuremath{q}\xspace}

}

\makeatletter
\ifcase \@ptsize \relax
  \newcommand{\miniscule}{\@setfontsize\miniscule{4}{5}}
\or
  \newcommand{\miniscule}{\@setfontsize\miniscule{5}{6}}
\or
  \newcommand{\miniscule}{\@setfontsize\miniscule{5}{6}}
\fi
\makeatother

\DeclareRobustCommand{\optbar}[1]{\shortstack{{\miniscule (\rule[.5ex]{1.25em}{.18mm})}
  \\ [-.7ex] $#1$}}

\def\quark     {{\ensuremath{\Pq}}\xspace}
\def\quarkbar  {{\ensuremath{\overline \quark}}\xspace}
\def\qqbar     {{\ensuremath{\quark\quarkbar}}\xspace}

\def\cquark    {{\ensuremath{\Pc}}\xspace}

\def\bquark    {{\ensuremath{\Pb}}\xspace}

\def\pion   {{\ensuremath{\Ppi}}\xspace}

\def\pip    {{\ensuremath{\pion^+}}\xspace}
\def\pim    {{\ensuremath{\pion^-}}\xspace}

\def\kaon    {{\ensuremath{\PK}}\xspace}
  \def\Kbar    {{\kern 0.2em\overline{\kern -0.2em \PK}{}}\xspace}

\def\KorKbar    {\kern 0.18em\optbar{\kern -0.18em K}{}\xspace}

\def\Kp      {{\ensuremath{\kaon^+}}\xspace}
\def\Km      {{\ensuremath{\kaon^-}}\xspace}

\def\Kstarz  {{\ensuremath{\kaon^{*0}}}\xspace}
\def\Kstarzb {{\ensuremath{\Kbar{}^{*0}}}\xspace}

  \def\Dbar    {{\kern 0.2em\overline{\kern -0.2em \PD}{}}\xspace}

\def\DorDbar    {\kern 0.18em\optbar{\kern -0.18em D}{}\xspace}

\def\Bbar    {{\ensuremath{\kern 0.18em\overline{\kern -0.18em \PB}{}}}\xspace}

\def\BorBbar    {\kern 0.18em\optbar{\kern -0.18em B}{}\xspace}

\def\jpsi     {{\ensuremath{{\PJ\mskip -3mu/\mskip -2mu\Ppsi\mskip 2mu}}}\xspace}

  \def\Y#1S{\ensuremath{\PUpsilon{(#1S)}}\xspace}

\def\proton      {{\ensuremath{\Pp}}\xspace}

\def\Deltares    {{\ensuremath{\PDelta}}\xspace}

\def\Lz          {{\ensuremath{\PLambda}}\xspace}
\def\Lbar        {{\ensuremath{\kern 0.1em\overline{\kern -0.1em\PLambda}}}\xspace}
\def\LorLbar    {\kern 0.18em\optbar{\kern -0.18em \PLambda}{}\xspace}

\def\Lb      {{\ensuremath{\Lz^0_\bquark}}\xspace}

\def\Lc      {{\ensuremath{\Lz^+_\cquark}}\xspace}

\def\Xicz    {{\ensuremath{\Xires^0_\cquark}}\xspace}
\def\Xicp    {{\ensuremath{\Xires^+_\cquark}}\xspace}

\newcommand{\decay}[2]{\ensuremath{#1\!\to #2}\xspace}         

\def\to                 {\ensuremath{\rightarrow}\xspace}

\def\CP                {{\ensuremath{C\!P}}\xspace}
\def\CPT               {{\ensuremath{C\!PT}}\xspace}

\def\AT#1     {\ensuremath{A_{\mathrm{T}}^{#1}}\xspace}           

\def\C#1      {\ensuremath{\mathcal{C}_{#1}}\xspace}                       
\def\Cp#1     {\ensuremath{\mathcal{C}_{#1}^{'}}\xspace}                    
\def\Ceff#1   {\ensuremath{\mathcal{C}_{#1}^{\mathrm{(eff)}}}\xspace}        
\def\Cpeff#1  {\ensuremath{\mathcal{C}_{#1}^{'\mathrm{(eff)}}}\xspace}       
\def\Ope#1    {\ensuremath{\mathcal{O}_{#1}}\xspace}                       
\def\Opep#1   {\ensuremath{\mathcal{O}_{#1}^{'}}\xspace}                    

\newcommand{\tev}{\ensuremath{\mathrm{\,Te\kern -0.1em V}}\xspace}
\newcommand{\gev}{\ensuremath{\mathrm{\,Ge\kern -0.1em V}}\xspace}
\newcommand{\mev}{\ensuremath{\mathrm{\,Me\kern -0.1em V}}\xspace}
\newcommand{\kev}{\ensuremath{\mathrm{\,ke\kern -0.1em V}}\xspace}
\newcommand{\ev}{\ensuremath{\mathrm{\,e\kern -0.1em V}}\xspace}
\newcommand{\gevc}{\ensuremath{{\mathrm{\,Ge\kern -0.1em V\!/}c}}\xspace}
\newcommand{\mevc}{\ensuremath{{\mathrm{\,Me\kern -0.1em V\!/}c}}\xspace}
\newcommand{\gevcc}{\ensuremath{{\mathrm{\,Ge\kern -0.1em V\!/}c^2}}\xspace}
\newcommand{\gevgevcccc}{\ensuremath{{\mathrm{\,Ge\kern -0.1em V^2\!/}c^4}}\xspace}
\newcommand{\mevcc}{\ensuremath{{\mathrm{\,Me\kern -0.1em V\!/}c^2}}\xspace}

\def\m    {\ensuremath{\mathrm{ \,m}}\xspace}

\def\cm   {\ensuremath{\mathrm{ \,cm}}\xspace}
\def\cma  {\ensuremath{{\mathrm{ \,cm}}^2}\xspace}

\def\mub{\ensuremath{{\mathrm{ \,\upmu b}}}\xspace}

\def\invfb   {\ensuremath{\mbox{\,fb}^{-1}}\xspace}

\def\sec  {\ensuremath{\mathrm{{\,s}}}\xspace}

\def\hz   {\ensuremath{{\mathrm{ \,Hz}}}\xspace}

\def\gsim{{~\raise.15em\hbox{$>$}\kern-.85em
          \lower.35em\hbox{$\sim$}~}\xspace}
\def\lsim{{~\raise.15em\hbox{$<$}\kern-.85em
          \lower.35em\hbox{$\sim$}~}\xspace}

\def\mrad{\ensuremath{\mathrm{ \,mrad}}\xspace}
\def\rad{\ensuremath{\mathrm{ \,rad}}\xspace}

\def\tell1  {TELL1\xspace}
\def\ukl1   {UKL1\xspace}

\newcommand{\eg}{\mbox{\itshape e.g.}\xspace}
\newcommand{\ie}{\mbox{\itshape i.e.}\xspace}

\usepackage{cite} 
\usepackage{mciteplus}
\newcommand{\br}{\ensuremath{\mathcal{B}} }

\newcommand{\thp}{\ensuremath{\theta' \xspace}}
\newcommand{\php}{\ensuremath{\phi' \xspace}}
\newcommand{\equationname}{Eq.}

\newcommand{\LctoLpi}{\ensuremath{\Lc\to\Lz(p\pim)\pip}\xspace}
\newcommand{\Lctopkpi}{\ensuremath{\Lc\to pK^-\pi^+}\xspace}

\newcommand{\Lctodeltak}{\ensuremath{\Lc\to \Deltares^{++}\Km}\xspace}

\def\pr          {{\ensuremath{p}}\xspace}
\def\PSigmap      {\ensuremath{\PSigma^+}\xspace}                 
\def\Lb      {{\ensuremath{\Lz^0_\bquark}}\xspace}

\def\Lc      {{\ensuremath{\Lz^+_\cquark}}\xspace}

\def\Xicz    {{\ensuremath{\PXi^0_\cquark}}\xspace}
\def\Xicp    {{\ensuremath{\PXi^+_\cquark}}\xspace}

\def\Xim    {\ensuremath{\PXi^-}\xspace}

\def\PH    {\ensuremath{H}\xspace}

\def\SH    {\ensuremath{S_{\PH}}\xspace}
\def\SL    {\ensuremath{S_{\mathrm L}}\xspace}
\def\SLz   {\ensuremath{S_{\Lz}}\xspace}
\def\SLzL   {\ensuremath{S_{\Lz\mathrm L}}\xspace}
\newcommand{\effCH}{\ensuremath{\varepsilon_{\rm CH}}\xspace}

\newcommand{\effDF}{\ensuremath{\varepsilon_{\rm DF}}}

\newcommand{\effdet}{\ensuremath{\varepsilon_{\rm det}}}
\newcommand{\effgeo}{\ensuremath{\varepsilon_{\rm geo}}\xspace}
\newcommand{\efftrigger}{\ensuremath{\varepsilon_{\rm trigger}}\xspace}
\newcommand{\efftrack}{\ensuremath{\varepsilon_{\rm track}}\xspace}

\def\invs{\ensuremath{{\mathrm{ \,s^{-1}}}}\xspace}
\def\invh{\ensuremath{{\mathrm{ \,h^{-1}}}}\xspace}

\usepackage{longtable} 

\begin{document}

\renewcommand{\thefootnote}{\fnsymbol{footnote}}
\setcounter{footnote}{1}

%
%
\begin{titlepage}
\vspace*{-1.5cm}
\vspace*{4.0cm}
{\normalfont\bfseries\boldmath\huge
\begin{center}
On the search for the electric dipole moment of strange and charm baryons at LHC
\end{center}
}
\vspace*{2.0cm}
\begin{center}
%
F.J. Botella$^1$,
L.M. Garcia Martin$^1$,
D. Marangotto$^2$,
F. Martinez Vidal$^1$\footnote{Corresponding author: fernando.martinez.vidal@cern.ch},\\
A. Merli$^2$,
N. Neri$^2$\footnote{Corresponding author: nicola.neri@cern.ch}, 
A. Oyanguren$^1$,
J. Ruiz Vidal$^1$
\bigskip\\
{\normalfont\itshape\footnotesize
$ ^1$ Instituto de F\'isica Corpuscular (IFIC), Universitat de Val\`encia-CSIC, Valencia, Spain\\ 
$ ^2$ INFN Sezione di Milano and Universit\`a di Milano, Milano, Italy\\
}
\end{center}
\vspace{\fill}
\begin{abstract}
Permanent electric dipole moments (EDMs) of fundamental particles provide powerful probes
for physics beyond the Standard Model. We propose to search for the EDM of strange
and charm baryons at LHC, extending the ongoing experimental program on the neutron, muon, atoms,
molecules and light nuclei. 
%
%
The EDM of strange \Lz baryons, selected from weak decays of charm
baryons produced in \proton\proton collisions at LHC,
can be determined by studying the spin precession in the magnetic field of the detector tracking system.
A test of \CPT symmetry can be performed by measuring the magnetic dipole moment of \Lz and \Lbar baryons.
For short-lived \Lc and \Xicp baryons, to be produced in a fixed-target experiment using the 7 \tev LHC beam
and channeled in a bent crystal, the spin precession is induced by the intense
electromagnetic field between crystal atomic planes.
The experimental layout based on the \lhcb detector and the expected sensitivities in the coming years are discussed.
%
%
%
%
%
%
%

\end{abstract}
\vspace*{0.5cm}
\begin{center}
\today
\vspace*{2.0cm}
\end{center}
\vspace{\fill}
\end{titlepage}
\pagestyle{empty} 
\newpage
\setcounter{page}{2}
\mbox{~}
%

\renewcommand{\thefootnote}{\arabic{footnote}}
\setcounter{footnote}{0}



\pagestyle{plain} 
\setcounter{page}{1}
\pagenumbering{arabic}


\section{Introduction}
\label{sec:intro}
%
%

The experimental searches for the electric dipole moment (EDM) of fundamental particles provide
powerful probes for physics beyond the Standard Model (SM).
The existence of permanent EDMs requires 
the violation of parity ($P$) and time reversal ($T$) symmetries and thus, relying on the validity of the \CPT theorem,
the violation of \CP symmetry.
Since EDM searches started in the fifties~\cite{Purcell:1950zz,Smith:1957ht},
there has been an intense experimental program,
leading to limits on the EDM of leptons~\cite{Baron:2013eja,Bennett:2008dy,Inami:2002ah},
neutron~\cite{Afach:2015sja}, heavy atoms~\cite{Griffith:2009zz},
proton (indirect from $^{199}$Hg)~\cite{Dmitriev:2003sc}, and \Lz baryon~\cite{Pondrom:1981gu}.
New experiments are ongoing and others are planned, including those based on storage rings
for muon~\cite{Grange:2015fou,Saito:2012zz}, proton and light nuclei~\cite{Anastassopoulos:2015ura,Pretz2015JEDI,Khriplovich:1998zq}.
Comprehensive reviews on EDM experiments can be found in 
Refs.~\cite{Engel:2013lsa,Fukuyama:2012np,Jungmann:2013sga,Pospelov:2005pr,Semertzidis:2011zz,Semertzidis:2016wtd,Onderwater2011}.

The amount of \CP violation in the weak interactions of quarks is not sufficient to explain
the observed 
imbalance between matter and antimatter in the Universe.
The SM Lagrangian of strong interactions contains a \CP-violating term proportional to the
QCD vacuum angle $\theta$; however, no \CP violation has been observed in the strong interactions.
A stringent upper bound, $\theta \lsim 10^{-10}$, is derived from the experimental limit
on the EDM of the neutron, 
$<3.0\times 10^{-26}~e\cm$ 
(90\% C.L.)~\cite{Afach:2015sja}.
This degree of tuning in the value of $\theta$ is known as the ``strong \CP'' problem.
Several solutions have been proposed, 
among which is the Peccei-Quinn mechanism~\cite{Peccei:1977hh,Weinberg:1977ma,Wilczek:1977pj}
that predicts the axion as a candidate for dark matter.

EDM searches of fundamental particles rely on the measurement of the spin precession angle
induced by the interaction with the electromagnetic field.
For unstable particles this is challenging since the 
precession has to take place before the decay. A solution to this problem requires large samples of
high energy 
polarized particles traversing an intense electromagnetic field.

In this work, we discuss the unique possibility to search for the EDM of the strange \Lz baryon
and of the charm \Lc and \Xicp baryons at LHC.
Using the experimental upper limit of the neutron EDM, the absolute value of the \Lz EDM is predicted
to be \mbox{$<4.4\times 10^{-26}~e\cm$~\cite{Guo:2012vf,Atwood:1992fb,Pich:1991fq,Borasoy:2000pq}}, while 
the indirect constraints
on the charm EDM are weaker, 
%
%
$\lsim 4.4\times 10^{-17}~e\cm$~\cite{Sala:2013osa}. 
Any experimental observation of an EDM 
would indicate a new source of \CP violation 
from physics beyond the SM.
The EDM of the long-lived \Lz baryon was measured to be
$<1.5 \times 10^{-16}~e\cm$ (95\% C.L.) in a fixed-target experiment
at Fermilab~\cite{Pondrom:1981gu}.
No experimental measurements exist for short-lived charm baryons
since negligibly small spin precession would be induced by
magnetic fields used in current particle detectors.

By studying the spin precession of polarized \Lz baryons, originated from weak charm baryon decays,
it is possible to extract the EDM.
We show that an improvement of the present limit 
of about two orders of magnitude is within reach of the \lhcb experiment.
The measurement of the magnetic dipole moment (MDM) of \Lz and \Lbar baryons 
would allow  a test of \CPT symmetry at per mille level.
A similar test has been performed for the proton~\cite{DiSciacca:2013hya},
electron~\cite{VanDyck:1987ay}, and muon~\cite{Bennett:2004pv}, and a new experiment for the proton is planned~\cite{Ulmer:2013rra}.

We propose to search for the EDM of short-lived charm baryons produced by interaction of
the 7\tev \lhc proton beam on a fixed target and channeled in a bent crystal in front of the \lhcb detector.
A sizeable spin precession angle for the short-lived \Lc and \Xicp baryons
would be possible by exploiting the intense electromagnetic field between crystal atomic planes.
The charm baryon decays can be reconstructed using the \lhcb detector. From one month dedicated runs,
sensitivities at the level of
$10^{-17}~e\cm$
can be reached.
This research would extend the physics program of the proposed
experiment~\cite{Baryshevsky:2016cul,Burmistrov:2194564} for the measurement of charm baryon MDMs.

\section{EDM experiment concept}
\label{sec:method}
%
%
%
%
%
%
%
%
%
%

The magnetic and electric dipole moment of a spin-1/2 particle is given (in Gaussian units) by
$\bm{\mu} = g \mu_B {\mathbf s}/2$ and \mbox{$\bm{ \delta} = d \mu_B {\mathbf s}/2$}, respectively,
where $\mathbf{s}$ is the spin-polarization vector\footnote{The spin-polarization vector
is defined such as 
$\mathbf s = 2 \langle \mathbf S \rangle / \hbar$, 
where $\mathbf S$ is the spin operator.
}
and $\mu_B=e \hbar / (2 m c)$ is the particle magneton, with $m$ its mass.
The $g$ and $d$ dimensionless factors 
are also referred to as the gyromagnetic and gyroelectric ratios.
The interaction of magnetic and electric dipole moments with external electromagnetic fields
causes the change of the particle spin direction.
The experimental setup to measure this effect relies 
on three main elements: i) a source of polarized particles whose direction 
and polarization degree are known;
ii) an intense electromagnetic field able to induce a sizable spin precession angle during 
the lifetime of the particle; iii) the detector to measure the final polarization vector
by analysing the angular distribution of the particle decays.

\subsection{\Lz and \Lbar case}
\label{sec:method_lambda}
A large amount of  \Lz baryons is produced directly from 
the \lhc\ \pr\pr collisions
via strong interactions. 
The initial polarization direction is perpendicular to the production plane, 
defined by the proton beam 
and \Lz momentum 
directions, due to parity conservation.
The level of polarization increases with the transverse momentum with respect to the beam direction.
Thus a significant initial polarization could be achieved 
by selecting events within 
specific kinematic regions~\cite{Heller:1978ty}.

In contrast, weak decays of heavy baryons (charm and beauty), mostly produced in the
forward/backward directions at \lhc, 
can induce large longitudinal polarization due to parity violation. 
%
%
%
%
%
%
For example, the decay of unpolarized \Lc baryons to the $\Lz\pip$ final state~\cite{Link:2005ft}, 
produces \Lz baryons with longitudinal polarization $\approx -90\%$, being the 
decay asymmetry parameter $\alpha_{\Lz\pip} = -0.91 \pm 0.15$~\cite{Olive:2016xmw}.
%
%
%
%
%
Another example is the $\Lb\to\Lz\jpsi$ decay where \Lz baryons are produced
almost 100\% longitudinally polarized~\cite{Aaij:2013oxa,Aad:2014iba}.
%
%

\begin{figure}[htb]
\centering
{ \includegraphics[width=0.31\linewidth]{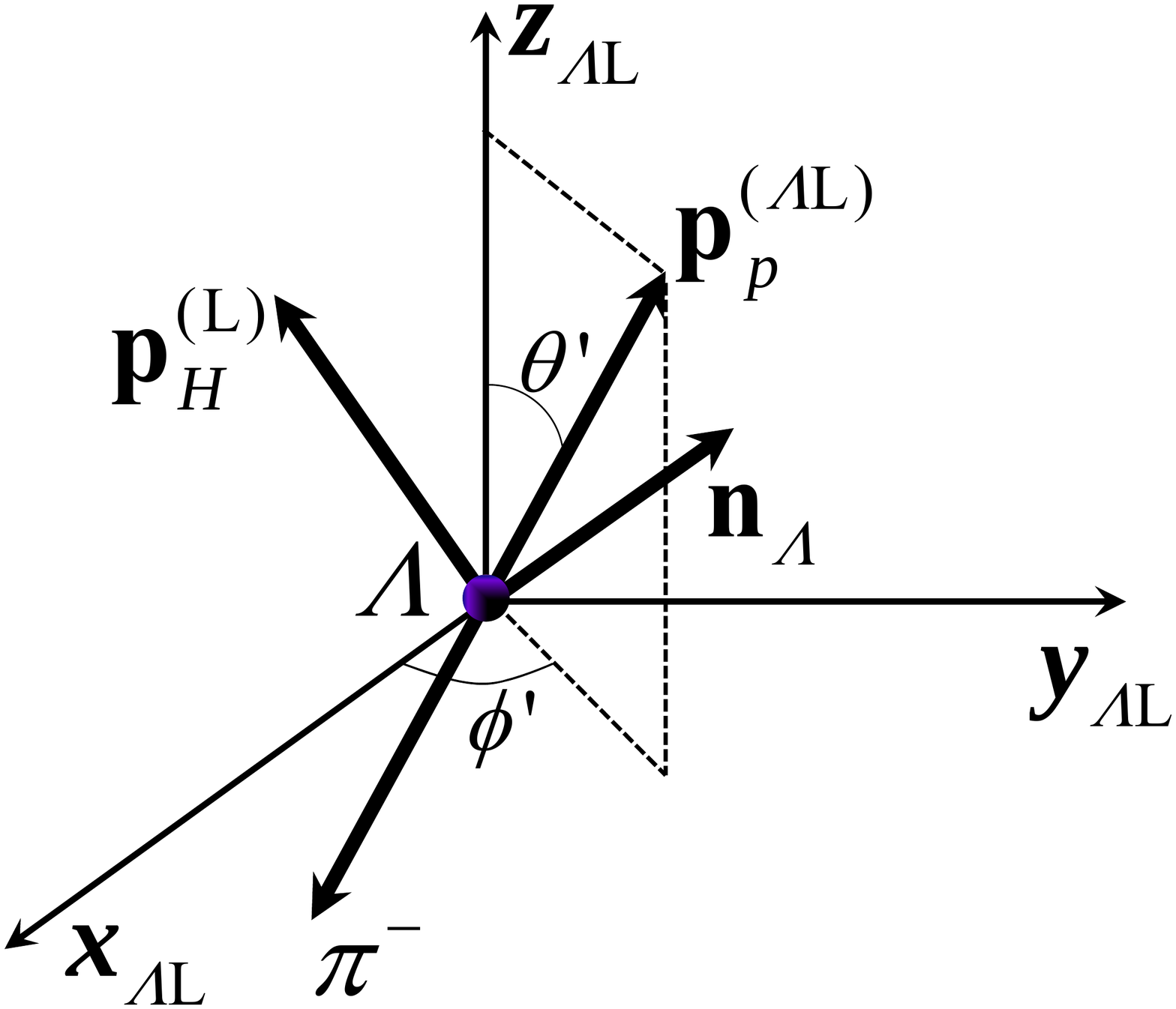} }
{ \includegraphics[width=0.31\linewidth]{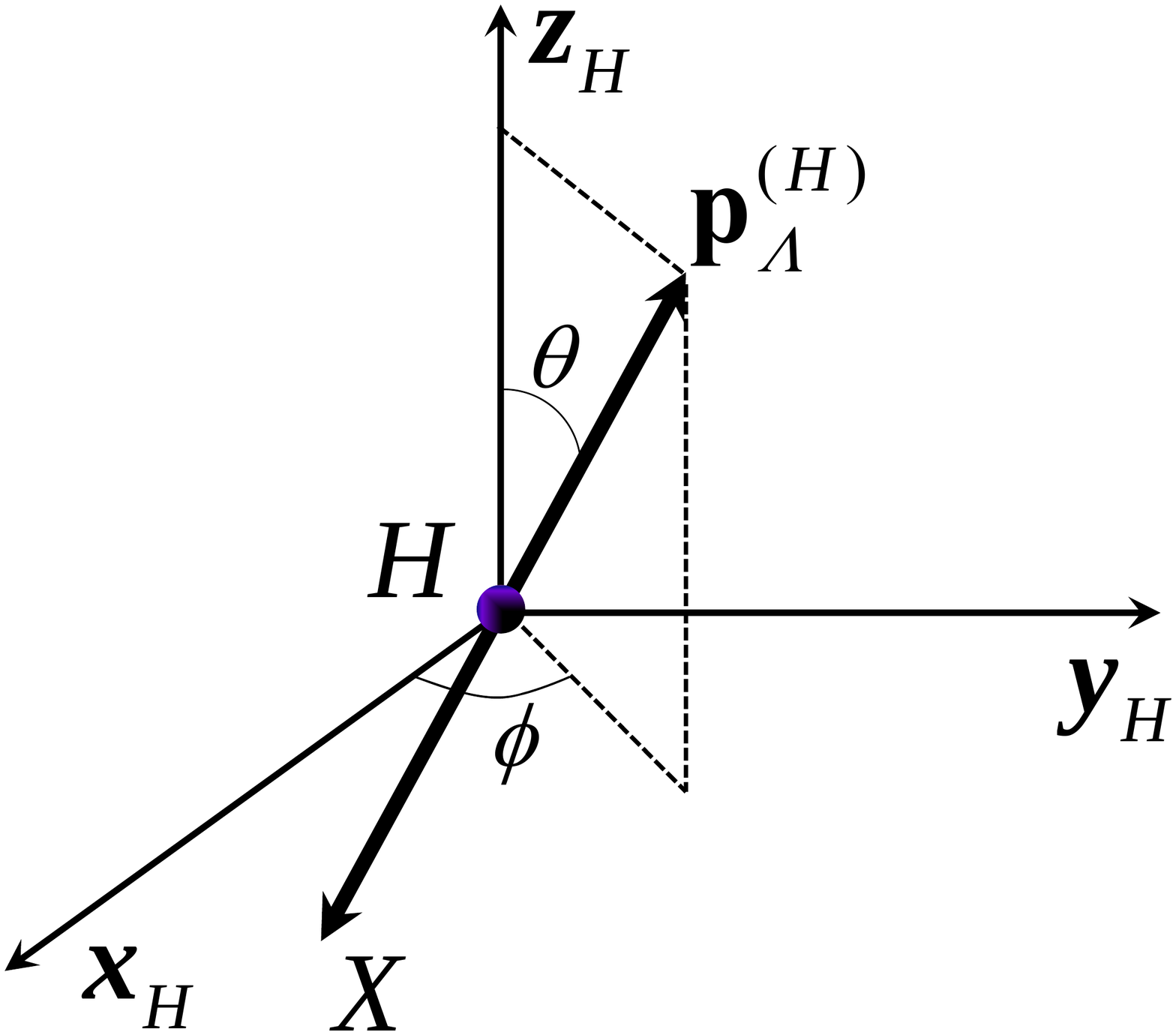} }
{ \includegraphics[width=0.31\linewidth]{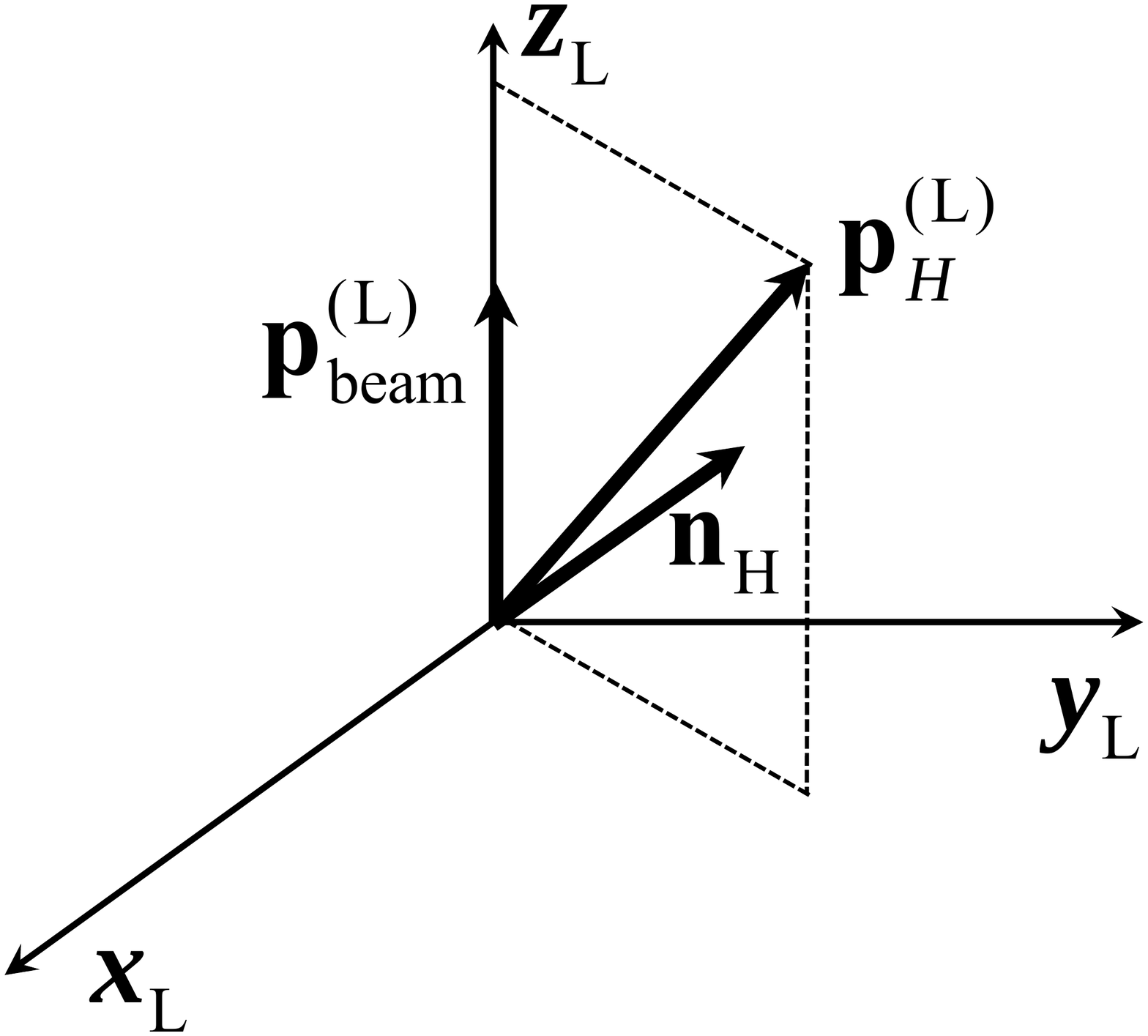} }
\caption{(Left) \Lz helicity frame (\SLzL), (Center) heavy baryon (\SH),
  and (Right) laboratory frame (\SL). The \Lz and proton angles, $(\thp,\php)$ and $(\theta,\phi)$
  are defined in the \SLzL and the \SH frames, respectively. 
The 
$z$
axis in \SLzL is defined by the \Lz momentum 
in \SL, 
and 
the $x$ axis is along the normal to the \Lz production plane, defined by the \Lz and \PH momenta in \SL frame.  
%
The 
$z$
axis in \SH is given by the heavy hadron momentum in \SL, 
and 
the $x$ axis
is 
parallel
to the normal to its production
plane.
%
The proton beam momentum 
is taken along the $z$ axis and the vertical direction 
by the $y$ axis in the \SL frame. 
}
\label{fig:frames}
\end{figure}

The spin-polarization vector $\mathbf s$ of an ensemble of \Lz baryons
can be analysed through the angular distribution of the $\Lz\to \pr \pim$ decay~\cite{Lee:1957qs,Richman1984},
%
\begin{equation}
\label{eq:AngDist}
\frac{dN}{d\Omega'} \propto 1 + \alpha \mathbf s \cdot \hat{\mathbf k} ~,
\end{equation}
where $\alpha = 0.642 \pm 0.013$~\cite{Olive:2016xmw} is the decay asymmetry parameter. 
The \CP invariance in the \Lz decay implies $\alpha = -\overline{\alpha}$, where $\overline\alpha$ is the decay parameter of 
the charge-conjugate decay.
The unit vector $\hat {\mathbf k} = (\sin\thp\cos\php, \sin\thp\sin\php, \cos\thp)$
indicates the momentum direction  of the proton in the \Lz 
helicity frame,
with $\Omega' = (\thp,\php)$ the corresponding solid angle, as illustrated in
(Left) Fig.~\ref{fig:frames}.
%
We can consider the \Lz momentum either in the 
heavy hadron helicity frame, \SH, shown in (Center) Fig.~\ref{fig:frames}, or 
in
the laboratory frame, \SL, defined in (Right) Fig.~\ref{fig:frames}.
This offers two possible options for the \Lz helicity frame, as seen from the \SH
or the \SL frames and referred to as \SLz or \SLzL, respectively, the latter sketched in (Left) Fig.~\ref{fig:frames}.
%
%

The dynamics of the spin vector in presence of external electromagnetic fields is given by the 
T-BMT
equation~\cite{Thomas:1926dy,Thomas:1927yu,Bargmann:1959gz} (see Appendix~\ref{app:A}). 
For a neutral particle in a magnetic field $\mathbf B$ in the laboratory with negligible 
field gradient effects, 
the general solution as a function of the \Lz flight length $l$ is described in
Sec.~\ref{app:lambda}.
%
For the particular case of \Lz and \PH baryons flying along the $z$ axis in \SL frame,
an initial longitudinal polarization $s_0$, \ie $\mathbf s_0=(0,0,s_0)$,
and $\mathbf B = (0,B_y,0)$, 
the solution is
\begin{equation}
\label{eq:sSimpleCase}
\mathbf s ~=~
\left\lbrace
\begin{array}{l}
s_x = - s_{0} \sin\Phi  \\
s_y = - s_{0} \dfrac{d \beta }{g} \sin\Phi \\
s_z =   s_{0} \cos\Phi \\
\end{array}
\right. 
\text{,~~~where~} {\Phi = \frac{D_y\mu_B}{\beta \hbar c} \sqrt{d^2 \beta^2 + g^2}
~~ \approx ~~ \frac{g D_y \mu_B}{\beta \hbar c} }~,
\end{equation}
with $D_y\equiv D_y(l) = \int_0^l B_y dl'$ the integrated magnetic field
along the \Lz flight path. 
The polarization vector precesses in the $xz$ plane, normal to the magnetic field, with 
the precession angle $\Phi$ proportional to the 
gyromagnetic factor of the particle.
The presence of an EDM introduces a non-zero
$s_y$ component perpendicular to the precession plane of 
the MDM, otherwise not present.
%
At LHCb, with a tracking dipole magnet providing an integrated field $D_y \approx \pm 4~\mathrm{T m}$~\cite{LHCb-DP-2014-002}, 
the maximum precession angle for particles traversing the entire magnetic field region yields $\Phi_{\rm max} \approx \pm \pi/4$, 
and allows to achieve about 70\% of the maximum $s_y$ component.
Moreover, 
a test of \CPT symmetry can be performed by comparing the $g$ and $-\bar g$ factors for \Lz and
\Lbar baryons, respectively, which precess in opposite directions as 
$g$ and $d$ change sign from particle to antiparticle.

Contrarily to the past fixed-target EDM experiments 
where the momentum direction in the laboratory frame was fixed and perpendicular to the magnetic field~\cite{Pondrom:1981gu,Schachinger:1978qs},
in this case the \Lz momentum varies being the particle produced from heavy baryon decays.
As a consequence, the polarization vector is not fixed to be perpendicular to the magnetic field
and the signature of the EDM becomes the variation of the $s_y$ component of the polarization vector
before and after the magnetic field.
To avoid the dilution introduced by the rotation of the \Lz production plane, the change of the
polarization has to be determined separately for ensembles of \Lz baryons with similar
initial polarization, selected according to the kinematics of the decay. In particular, 
the projection of the \Lz trajectory in the $xy$ plane in \SL at the $z$ position of the \PH production vertex can be used 
to select events with similar polarization,
as discussed in Sec.~\ref{app:lambda_rotations}.

%
%
%
%

\subsection{\Lc and \Xicp case}
\label{sec:method_lambdac}
%
%
%
%
%
%
%
The \Lc and the \Xicp baryon EDM can be extracted by measuring the precession of the
polarization vector of channeled particles in a bent crystal.
There, a positively-charged particle channeled between atomic planes moves along a curved path under
the action of the intense electric field between crystal planes. In the instantaneous rest frame of the particle
the electromagnetic field causes the spin rotation.
The signature of the EDM
is a polarization component perpendicular to the initial baryon
momentum and polarization vector, otherwise not present, similarly to the case of
the \Lz baryon.

The phenomenon of spin precession of positively-charged particles channeled in a bent crystal
was firstly observed by the E761 collaboration that measured the MDM
of the strange \PSigmap baryon~\cite{Chen:1992wx}.
The possibility to measure the MDM of short-lived charm baryons using channeling in bent crystals,
in the momentum range of hundreds of \gevc, is discussed in Ref.~\cite{Baublis:1994ku,Samsonov:1996ah}.
The feasibility of the measurement at \lhc\ energies is studied in Ref.~\cite{Baryshevsky:2016cul}
and offers clear advantages with respect to lower beam energies since the estimated number
of produced charm baryons that are channeled into the crystal is proportional to $\gamma^{3/2}$ where $\gamma$ is the Lorentz factor of the particles. 
 
\begin{figure}[H]
\centering
{ \includegraphics[width=0.35\linewidth]{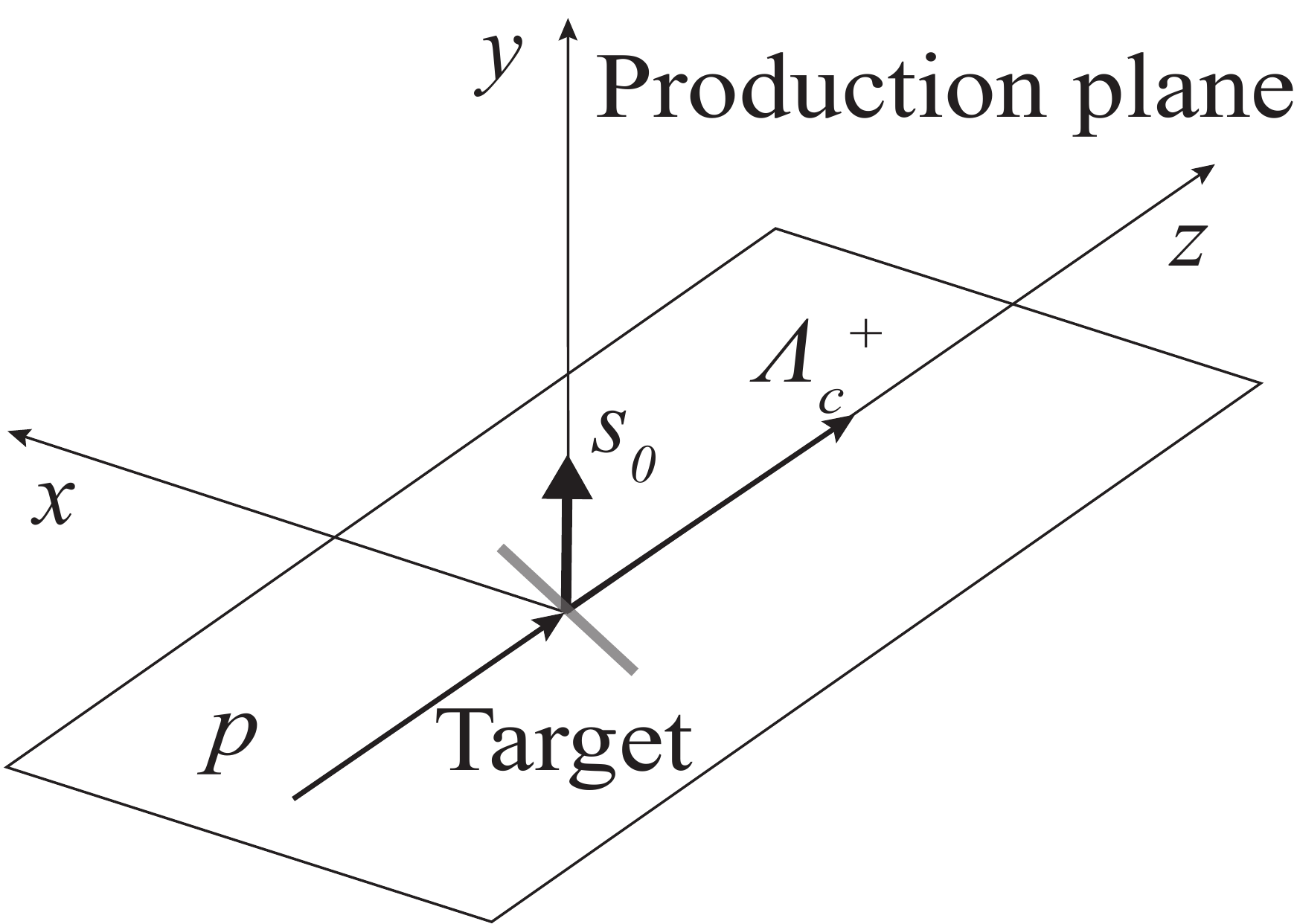} }
\hspace{1cm}    
{ \includegraphics[width=0.55\linewidth]{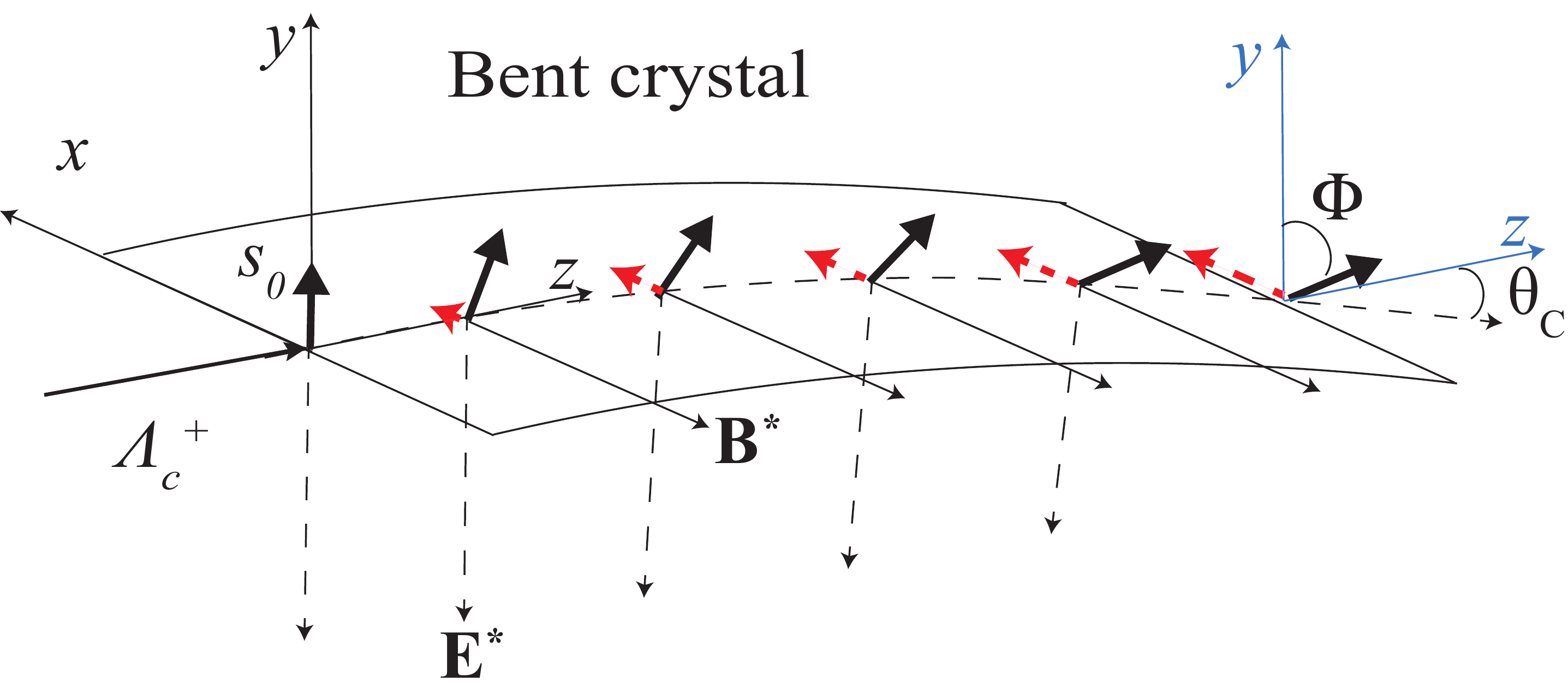} }
\caption{(Left) Production plane of the \Lc baryon defined by the proton and the \Lc momenta.
  The initial polarization vector $\mathbf s_0$ is perpendicular to the production plane,
  along the $y$ axis, due to parity conservation in strong interactions.
  (Right) Deflection of the baryon trajectory and
  spin precession in the $yz$ and $xy$ plane induced by the MDM and the EDM, respectively.
  The red (dashed) arrows indicate the (magnified) $s_x$ spin component
  proportional to the particle EDM. $\Phi$ is the MDM precession angle and $\theta_C$ is
 the crystal bending angle.}
\label{fig:Lc_ProdBending}
\end{figure}

Charm baryons produced by interaction of protons on a fixed target, \eg tungsten target,
are polarized perpendicularly to the production
plane due to parity conservation in strong interactions~\cite{Jacob:1959at}.
The production plane $xz$, shown in (Left) Fig.~\ref{fig:Lc_ProdBending}, is determined by the
proton and the charm baryon momenta; the latter defines the $z$ axis. The initial polarization
vector $\mathbf s_0 =(0,s_0,0)$ is perpendicular to the production plane, along the $y$ axis.
To induce spin rotation the crystal is bent in the $yz$ plane.

The intense electric field $\mathbf E$ between the crystal planes which deflects positively-charged particles, 
transforms into a strong electromagnetic field
$\mathbf{E^*}\approx\gamma \mathbf{E}$,
$\mathbf{B^*}\approx-\gamma \bm\beta\times\mathbf{E} /c$ in the particle rest frame
%
and induces the spin precession, as it is described in detail in Refs.~\cite{Baryshevsky:2015zba,Kim:1982ry} and
illustrated in (Right) Fig.~\ref{fig:Lc_ProdBending}.
The crystal bending angle is defined as $\theta_C=L/\rho_0$, where $L$ is the circular arc of the crystal
and $\rho_0$ the curvature radius.
The precession angle $\Phi$ is defined as the angle between the polarization vector and the
$y$ axis, as shown in (Right) Fig.~\ref{fig:Lc_ProdBending}.
In the limit of large boost with Lorentz factor $\gamma \gg 1$, the precession angle 
in the $yz$ plane
induced by the MDM is~\cite{Lyuboshits:1979qw}
\begin{equation}
\Phi \approx \frac{g-2}{2}\gamma \theta_C,
\label{eq:MDM_angle}
\end{equation}
where $g$ is the gyromagnetic factor.
%
%
%

In presence of a non-zero EDM, the spin precession is no longer confined to the $yz$ plane,
originating a $s_x$ component proportional
to the particle EDM represented by the red (dashed) arrows in (Right) Fig.~\ref{fig:Lc_ProdBending}.
The integration of the equation of motion in presence of EDM
is described in Appendix~\ref{app:A}, as well as the approximations used to solve the
equations analytically. The polarization vector, after channeling through the crystal is 
\begin{equation}
\mathbf s ~=~
\left\lbrace
\begin{array}{l}
s_{x} \approx   s_0 \dfrac{d}{g-2}  (\cos{\Phi}-1)  \\
s_{y} \approx   s_{0} \cos\Phi \\
s_{z} \approx   s_{0} \sin\Phi
\end{array}
\right. ,
\label{eq:EDM_LcPol}
\end{equation}
where $\Phi$ is given by Eq.~(\ref{eq:MDM_angle}).
%
%
%
%
%
The polarization can be determined, as in the case of the \Lz EDM
described in Sec.~\ref{sec:method_lambda},
by studying the angular distribution of the final state particles. 
The angular distribution for non-channeled particles allows to determine
the initial polarization along the $y$ axis,
which compared to the final polarization 
allows to extract the gyromagnetic and gyroelectric factors.
The same method applies to both \Lc and \Xicp baryons.

For \Lc decaying to two-body
final states such as $p\Kstarz$, $\Deltares^{++}\pim$, $\Lz(1520)\pip$ and $\Lz\pip$, the angular distribution 
is described by Eq.~(\ref{eq:AngDist}),
%
%
where $\alpha$ is a parity violating coefficient depending on the final state,
$\hat{\mathbf k}$ the 
direction of the final state baryon
in the \Lc helicity frame,
and ${\bf s}$ the \Lc polarization vector. 
In the case of the $\Lc\rightarrow \Lz\pim$ decay, the $\alpha$ parameter is measured to be
$\alpha_{\Lz\pim} = -0.91\pm0.15$~\cite{Olive:2016xmw}.
%
For other \Lc decays no measurements are available but an effective $\alpha$ parameter
can be  calculated from a Dalitz plot analysis of $\Lc\to\proton\Km\pip$
decays~\cite{Aitala:1999uq}, 
as discussed in Appendix~\ref{app:B} and summarized in Table~\ref{tab:alphas}.
Eventually, a Dalitz plot analysis would provide the ultimate 
sensitivity to the EDM measurement.
The initial polarization $s_0$ of \Lc particles produced from the interaction of 7\tev protons on
a fixed target has not been measured. However, a measurement of \Lc polarization from
40-70 \mevc neutron on carbon target gives $s_0=0.5\pm0.2$~\cite{Szwed:1981rr},
and a measurement from interaction of 230 \mevc~\pim on copper target 
yields
$s_0=-0.65^{+0.22}_{-0.18}$~\cite{Jezabek:1992ke}. 


%
\section{Sensitivity studies}
\label{sec:sensitivity}

\subsection{\Lz and \Lbar case}
\label{sec:sensivity_lambda}
%



To identify the 
most copious
\Lz production channels 
from heavy baryons, 
we consider decays
containing
only charged particles in the final state, with at least one originated from the heavy baryon decay vertex.
No other long-living particles besides the \Lz baryon, except an intermediate \Xim baryon
decaying into the $\Lz\pim$ final state,  are considered.
These conditions are required to reconstruct the production and the decay vertex of 
the \Lz particle and eventually exploit this information in the event reconstruction.
%
%
%
The number of \Lz particles produced can be estimated as 
\begin{equation}
N_\Lz=2 \mathcal{L} \sigma_\qqbar f(\quark \to \PH)\br(\PH \to \Lz X')\br(\Lz\to\pr\pim)\br(X'\to\mathrm{charged}) , 
\end{equation}
where $\mathcal{L}$ is the total integrated luminosity, $\sigma_\qqbar$ ($\quark=\cquark,\bquark$) are the heavy quark 
production cross sections from \pr\pr collisions at $\sqrt{s}=13$\tev~\cite{Aaij:2015bpa,FONLLWEB,Aaij:2010gn,Aaij:2015rla},
and $f$ is the fragmentation fraction into the heavy 
baryon \PH~\cite{Lisovyi:2015uqa,Gladilin:2014tba,Amhis:2014hma,Galanti:2015pqa}.
%
%
%
All branching fractions \br are taken from Ref.~\cite{Olive:2016xmw}, and where
they are given relative to other decays all the known decay modes are assumed to sum
the total width.
In Table~\ref{tab:LambdaChannels} the dominant production channels and the estimated
 yields are summarised.
Overall, there are about $1.5\times 10^{11}$ \Lz baryons per \invfb produced directly from heavy baryon decays 
(referred hereafter as short-lived, or SL events), 
and $3.8\times 10^{11}$ from charm baryons decaying through an intermediate \Xim particle (long-lived, or LL events).
The yield of \Lz baryons experimentally available can then be evaluated as
$N_\Lz^{\rm reco} = \epsilon_{\rm geo} \epsilon_{\rm trigger} \epsilon_{\rm reco} N_\Lz$,
where $\epsilon_{\rm geo}$, $\epsilon_{\rm trigger}$ and $\epsilon_{\rm reco}$ are
the geometric, trigger and reconstruction efficiencies of the detector system.
 
\begin{table}[htb]
\centering
\caption{Dominant \Lz production mechanisms from heavy baryon decays and estimated yields
  produced per \invfb at $\sqrt{s}=13$\tev,
shown separately for SL and LL topologies.
The \Lz baryons from \Xim decays, produced promptly in 
the \pr\pr collisions, are given in terms of the unmeasured production cross section. 
}
\label{tab:LambdaChannels}
\renewcommand{\arraystretch}{1.1}
\begin{tabular}{lc lc}
\toprule
SL events &  $N_{\Lz}/\invfb~(\times 10^{10})$  & LL events, $\Xim\to\Lz\pim$ &  $N_{\Lz}/\invfb~(\times 10^{10})$  \\ 
\midrule
$\Xicz\to\Lz\Km\pip$       & 7.7 & $\Xicz\to\Xim\pip\pip\pim$ & 23.6 \\
$\Lc\to\Lz\pip\pip\pim$    & 3.3 & $\Xicz\to\Xim\pip$         & 7.1 \\
$\Xicp\to\Lz\Km\pip\pip$   & 2.0 & $\Xicp\to\Xim\pip\pip$     & 6.1 \\
$\Lc\to\Lz\pip$            & 1.3 & $\Lc\to\Xim\Kp\pip$        & 0.6 \\
$\Xicz\to\Lz\Kp\Km$ (no $\phi$)  & 0.2 & $\Xicz\to\Xim\Kp$              & 0.2 \\
$\Xicz\to\Lz\phi(\Kp\Km)$  & 0.1 & Prompt $\Xim$              & $0.13\times\sigma_{\pr\pr\to\Xim}~[\mu \rm b]$ \\
\bottomrule
\end{tabular}
\end{table}

\begin{figure}[htb]
\centering
{ \includegraphics[width=0.48\linewidth]{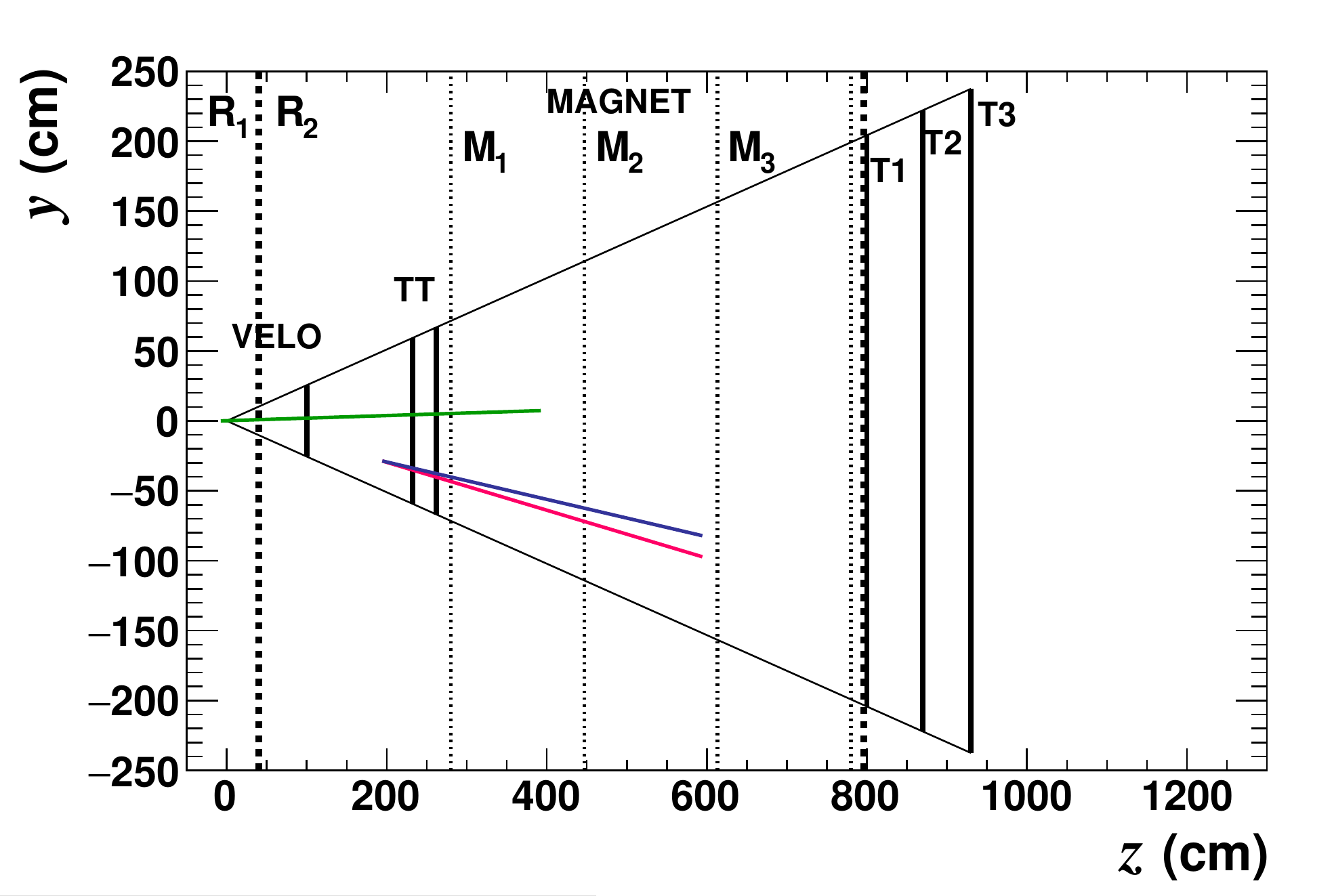} }
{ \includegraphics[width=0.48\linewidth]{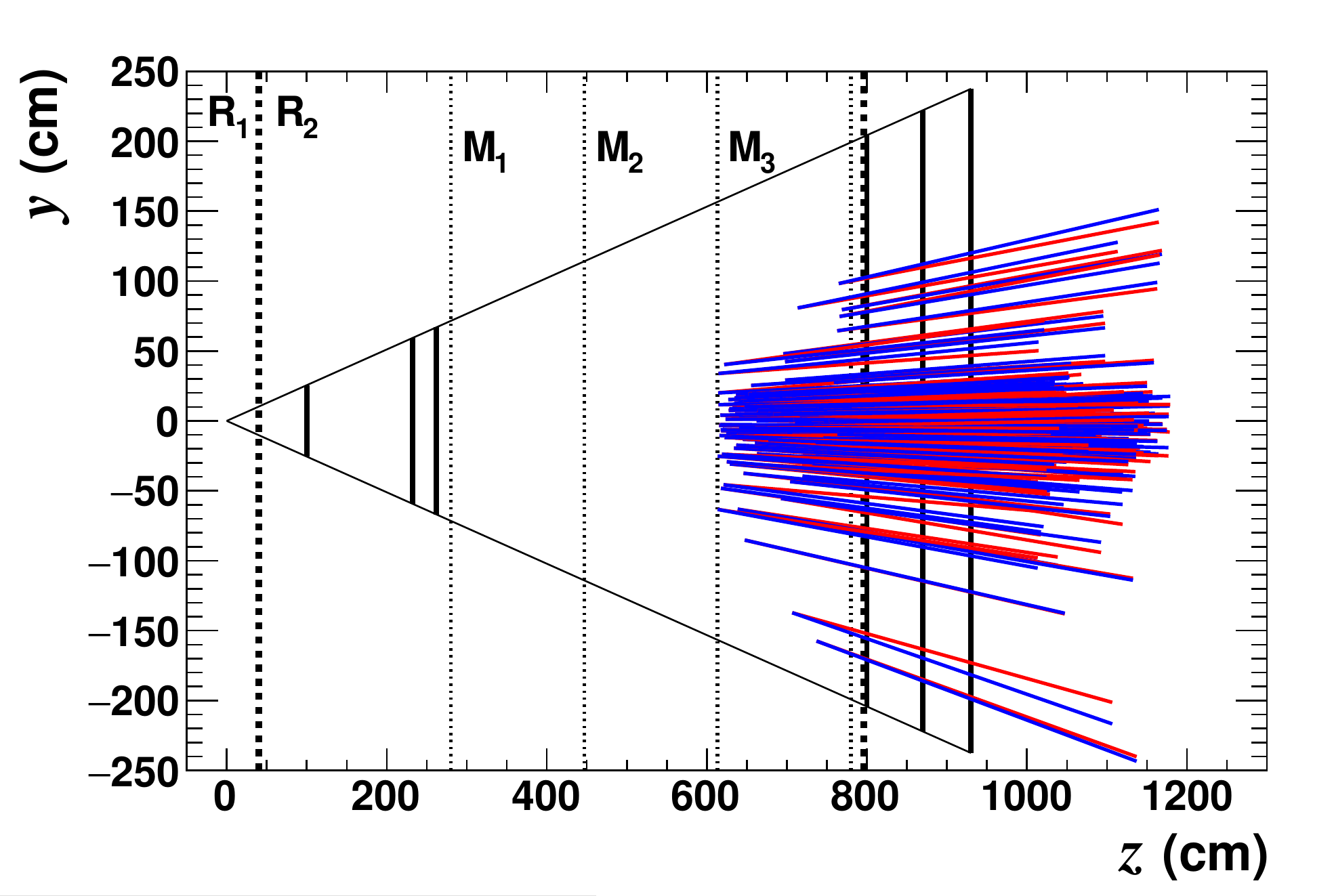} }
\caption{(Left) Sketch of the simplified geometry of the LHCb tracking system in the $yz$ plane.
The crosswise lines represent the angular acceptance. The tracking layers and the limits
of the R$_1$ and R$_2$ regions are shown as solid and dotted thick lines, respectively.
The magnet is divided in three regions by thin dotted lines. 
A simulated $\Lc \to\Lz (p \pim)\pip$ decay with corresponding $\pip$ (green), $\pim$ (blue) and $p$ (red) tracks is overlaid. 
(Right) Decay products from \Lz baryons decaying in the
last region of the magnet, M$_3$. 
}
\label{fig:detectorDiagram}
\end{figure}

The geometric efficiency for SL topology has been estimated using a 
Monte Carlo simulation of $\pr\pr$ collisions at $\sqrt{s}=13$\tev and the decay of heavy hadrons, 
using Pythia~\cite{Sjostrand:2006za} and EvtGen~\cite{Lange:2001uf} standalone toolkits, together with a 
simplified geometrical model of the LHCb detector~\cite{LHCb-DP-2014-002}.
Tracking devices upstream of the dipole magnet (VErtex LOcator and Tracker Turicensis)
and downstream the magnet (T stations) 
are modelled to have rectangular shape. The height and width of the tracking layers
along the beam axis are determined by the detector angular acceptance, between
10 \mrad and 250 \mrad (300 \mrad) in the vertical (horizontal) direction,
as illustrated in (Left) Fig.~\ref{fig:detectorDiagram}.
%
%
%
Particle trajectories are approximated by straight lines defined by the momentum directions.

Table~\ref{tab:LambdaGeoEfficiencies} summarizes the geometric efficiencies for \Lz baryons decaying in different regions of the
detector volume, for three different SL topologies. Region R$_1$ is defined such that the $z$ position of the \Lz decay vertex 
is in the range [0-40]\cm from the collision point
and the decay products are within the detector acceptance.
Events in the R$_2$ region have a \Lz decay $z$ position in the range [40-800]\cm.
Charged particles produced together with the \Lz baryon are required to be within the
VELO and T1-T3, or the VELO and TT acceptances, to insure a precise reconstruction of
the \Lz origin vertex.
Events in the R$_1$ region provide the measurement of the initial \Lz polarization vector;
events in the R$_2$ region allow to determine the polarization as a function of the \Lz decay
length in the magnetic field region. Among the latter, \Lz baryons decaying towards the end
of the magnet (M$_3$ region in Table~\ref{tab:LambdaGeoEfficiencies}) provide most of
the sensitivity to the EDM and MDM.
These events are sketched in (Right) Fig.~\ref{fig:detectorDiagram}.
The total geometric efficiency for R$_1$ and R$_2$ regions is about 16\%, with small differences among SL topologies, 
and about $2.4\times 10^{10}$ \Lz baryons per \invfb can be
reconstructed.
%

\begin{table}[htb]
\centering
\caption{Geometric efficiencies (in \%) for \Lz baryons decaying in different regions
of the \lhcb detector, for several charm baryon decays produced at $\sqrt{s}=13$\tev.}
\label{tab:LambdaGeoEfficiencies}
\renewcommand{\arraystretch}{1.1}
\begin{tabular}{l ccccc}
\toprule
  Region  &  R$_1$  & R$_2$ &  M$_1$ & M$_2$ & M$_3$ \\ 
\Lz decay vertex $z$ position (cm)   &  [0-40] & [40-800] & [280-450] & [450-610] &[610-780]  \\  
\midrule
$\Lc\to\Lz\pip\pip\pim$    & 4.7 & 10.5 & 1.3 & 0.7 & 0.3  \\
$\Xicz\to\Lz\Km\pip$       & 5.2 & 12.2 & 1.7 & 1.0 & 0.6  \\
$\Xicp\to\Lz\Km\pip\pip$   & 5.3 & 11.9 & 1.6 & 0.9 & 0.4  \\
\toprule
\end{tabular}
\end{table}

\begin{figure}[htb]
\centering
{ \includegraphics[width=0.48\linewidth]{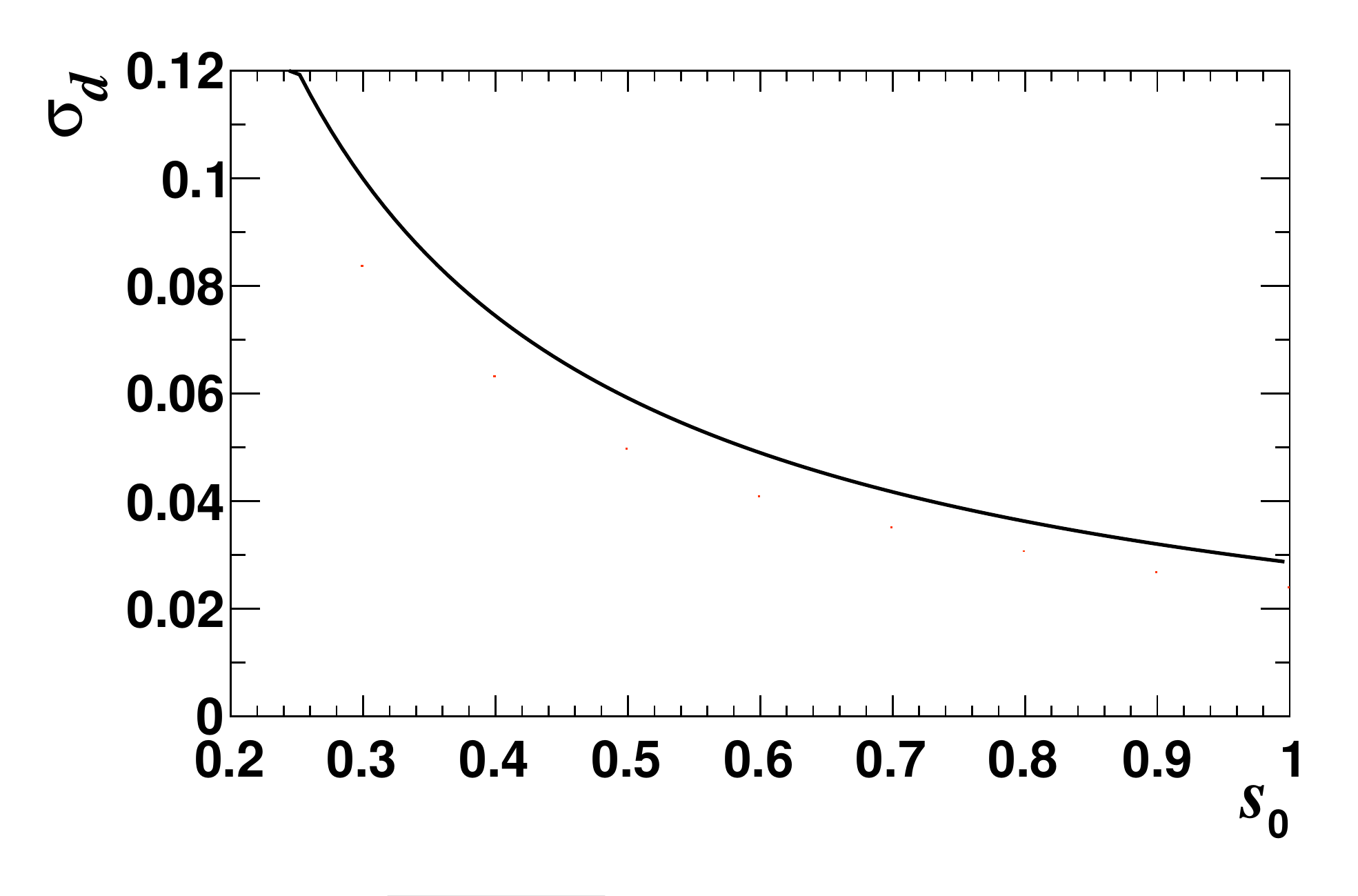} }
{ \includegraphics[width=0.48\linewidth]{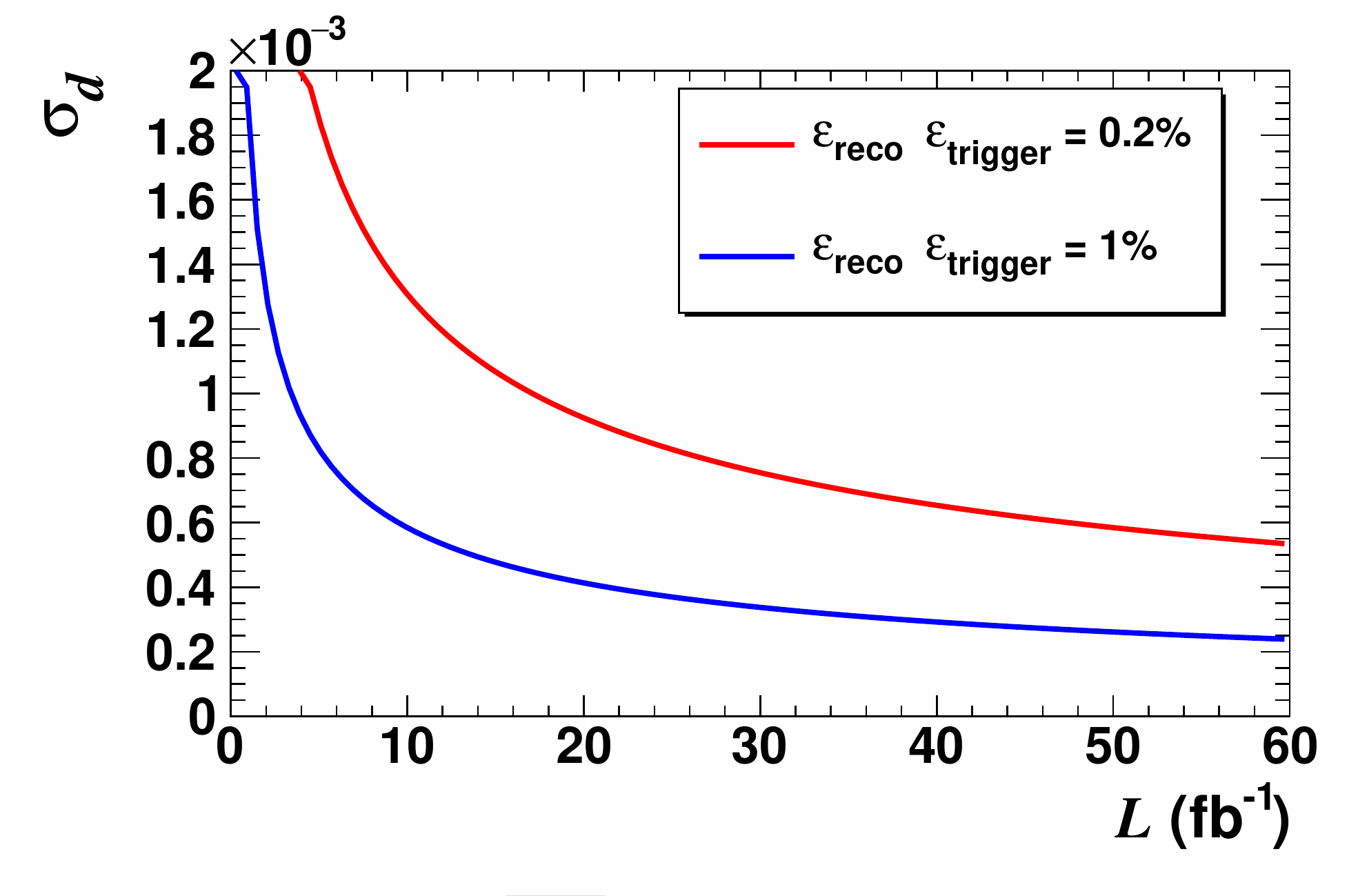} }
\caption{(Left) Dependence of the $d$ uncertainty with the initial polarization for $N_\Lz^{\rm reco}=10^6$ events, and (Right) as a function 
of the integrated luminosity
assuming reconstruction efficiency of  0.2\% and 1\%.}
\label{fig:Lambda_sensitivity}
\end{figure}

To assess the EDM sensitivity, pseudo-experiments have been generated using a simplified detector geometry
that includes an approximate \lhcb magnetic field mapping~\cite{LHCb-DP-2014-002,Hicheur:2007jfk}. 
The angular distribution and spin dynamics have been simulated 
using Eq.~(\ref{eq:AngDist}) and the general solution as a function of the \Lz flight length 
described in Sec.~\ref{app:lambda}, respectively.
%
For this study initial polarization vector $\mathbf s_0 = (0,0,s_0)$, 
with $s_0$ varying between 20\% and 100\%, and factors $g=-1.458$~\cite{Olive:2016xmw} and $d=0$,
were used. Each generated sample was adjusted using an unbinned maximum likelihood fitting method with $d$, 
$g$ and $\mathbf s_0$ (or $\alpha\mathbf s_0$) as free parameters. The $d$-factor uncertainty 
scales with the number of events $N_\Lz^{\rm reco}$ and the initial longitudinal polarization $s_0$ as
$\sigma_d \propto 1/(s_0 \sqrt{N_\Lz^{\rm reco}} )$. 
The sensitivity saturates at large values of $s_0$, as shown in (Left) Fig.~\ref{fig:Lambda_sensitivity},
and it partially relaxes 
the requirements on the initial polarizations.
Similarly, (Right) Fig.~\ref{fig:Lambda_sensitivity} shows the expected sensitivity on the EDM as a function 
of the integrated luminosity, summing together SL and LL events, assuming global trigger
and reconstruction efficiency $\epsilon_{\rm trigger} \epsilon_{\rm reco}$ 
of 1\% (improved \lhcb software-based trigger and tracking for the upgrade detector~\cite{LHCb-TDR-016,LHCb-TDR-015}) 
and 0.2\% (current detector~\cite{LHCb-DP-2014-002}), where the efficiency estimates are based on a educated guess.
An equivalent sensitivity is obtained for the gyromagnetic factor.
Therefore, with 8~\invfb a sensitivity $\sigma_d \approx 1.5\times 10^{-3}$ could be achieved (current detector), 
to be compared to the present limit, $1.7\times 10^{-2}$~\cite{Pondrom:1981gu}. 
With 50~\invfb (upgraded detector) the sensitivity on the gyroelectric factor can reach $\approx 3\times 10^{-4}$. 

The reconstruction of long-lived \Lz baryons decaying inside and after the magnet
represents a challenge for the \lhcb experiment,
introducing significant backgrounds and a limited resolution on the measurement of the 
\Lz momentum and decay point. 
Events can be reconstructed by exploiting the kinematics
of exclusive decays and the determination of the production and the decay vertex of the \Lz.
According to simulation studies even with relatively poor resolutions, the EDM and MDM measurements do not degrade significantly.

\subsection{\Lc and \Xicp case}
\label{sec:sensivity_lambdac}
We propose to search for charm baryon EDMs in a dedicated fixed-target experiment at the LHC to be
installed in front of the \lhcb detector, as close as possible to the VELO detector. The target should
be attached to the crystal to maximize the yield of short-lived charm baryons to be channeled.
The rate of \Lc baryons produced with 7\tev protons on a fixed target can be estimated as
\begin{equation}
\frac{dN_\Lc}{dt} = \frac{F}{A}\sigma(\proton\proton\rightarrow\Lc X) N_T ,
\end{equation}
where $F$ is the proton rate, $A$ the beam transverse area, $N_T$ the number of target nucleons, and
$\sigma(\proton\proton\rightarrow\Lc X)$ is the cross-section for \Lc production in \proton\proton interactions at
 $\sqrt{s}=114.6\gev$ center-of-mass energy.
%
The number of target nucleons is $N_T=N_A\rho A T A_N/A_T $,
where $N_A$ is the Avogadro number, $\rho$ ($T$) is the target density (thickness),
and $A_T$ ($A_N$) is the atomic mass (atomic mass number).
The rate of \Lc particles channeled in the bent crystal and reconstructed in the \lhcb detector is estimated as
\begin{equation}
\label{eq:NLcReco}
  \frac{dN_\Lc^{\rm reco}}{dt} = \frac{dN_\Lc}{dt} \br(\Lc\to f)\effCH\effDF(\Lc)\effdet
\end{equation}
where each quantity and the corresponding estimated value is defined in Table~\ref{tab:quantities}.
\begin{table}[!h]
\caption{Definitions and estimated values of the relevant quantities for charm
   baryon EDM and MDM sensitivity studies, for a tungsten (W) target.\label{tab:quantities}}
\makebox[\textwidth]
{\begin{tabular}{lccc}
\toprule
Definition & Quantity & Value & Unit\\
\midrule
Proton flux on target & $F$ & $5\times 10^{8}$ & proton/\sec\\
\midrule
Avogadro number & $N_A$ & $6.022\times 10^{23}$ & atoms/mol\\
Target density (W) & $\rho$ & $19.25$ & g/$\cm^3$\\
Target thickness & $T$ & 0.5 & \cm\\
Atomic mass (W) & $A_T$ & 183.84 & g/mol\\
Atomic mass number (W) & $A_N$ & 183.84 &\\
\midrule
\proton\proton cross-section to \Lc & $\sigma(\proton\proton\to\Lc X)$ & 18.2 & \mub \\
Branching fraction~\cite{Olive:2016xmw} & \br(\Lctodeltak) & $1.09\%$ & \\
& \br(\LctoLpi) & $0.83\%$ & \\
\Lc boost & $\gamma$ & $10^{3}$\\
\midrule
Crystal length & $L$ & 10 & \cm\\
Crystal radius & $\rho_0$ & 10 & \m\\
\midrule
Channeling efficiency & \effCH & $10^{-3}$\\
Decay flight efficiency & \effDF(\Lc) & $19\%$\\
& \effDF(\Xicp) & $47\%$\\
Detector efficiency & \effdet(\Lctopkpi) & $5.4\%$\\
& \effdet(\LctoLpi) & $10^{-3}$\\
\midrule
\Lc polarization & $s_0$ &  0.6\\
$\alpha$ parameter & $\alpha_{\Lz\pip}$ &  $-0.91$\\
& $\alpha_{\Deltares^{++}\Km}$ & $-0.67$\\
MDM anomaly & $(g-2)/2$ & 0.3\\
\bottomrule
\end{tabular}}
\end{table}
A 6.5\tev proton beam was extracted from the LHC beam halo by channeling protons in
bent crystals~\cite{Scandale:2016krl}. A beam with intensity of $5\times 10^8~\text{proton/\sec}$,
to be directed on a fixed target, is attainable with this technique~\cite{Lansberg:2012wj}.
An alternative experimental setup to be considered is 
a target-crystal system
positioned in the vacuum pipe of the LHC where collisions with protons of the beam halo can
be reached at comparable rates.
Both solutions should be studied very accurately to be compliant with machine protection
and safety requirements. Recent results from the UA9 collaboration~\cite{Scandale:2016krl}, relative to
crystal collimation tests, demonstrated that a similar setup is technically viable and
can be installed successfully in the LHC. Fixed-target collision events can be recorded in short dedicated runs or
in parallel to the \proton\proton data taking, if the background caused by the insertion
of a fixed target in the beam halo is negligible with respect to \proton\proton collisions. Both solutions
 have to be studied in detail using ad-hoc simulations.
%
%

The \Lc cross section can be estimated from the total charm production cross section measured by the PHENIX experiment in proton-proton collisions at $\sqrt{s} = 200\gev$~\cite{Adare:2006hc},
$\sigma_{c\overline{c}} = \left(567 \pm 57_{\rm stat.} \pm 193_{\rm syst.}\right) \mub$,
rescaled to $\sqrt{s} = 114.6 \gev$ assuming a linear dependence 
on $\sqrt{s}$. By applying the \Lc fragmentation function used in Ref. \cite{Adare:2006hc},
$\sigma_{\Lc}/\sigma_{c\overline{c}}\approx 5.6\%$, compatible with theoretical predictions \cite{Kniehl:2005de}, the \Lc cross section is
$\sigma_{\Lc} \approx 18.2 \mub$.

The channeling efficiency in silicon crystals, including both channeling angular acceptance and dechanneling effects, is estimated to be $\effCH\approx 10^{-3}$~\cite{Biryukov1997}, while
%
%
the fraction of \Lc baryons decaying after the crystal is $\effDF(\Lc)\approx 19\%$, for 
$\gamma = 1000$ and 10\cm crystal length.
%
The geometrical acceptance for $\Lc\to\proton\Km\pip$ decaying into the \lhcb detector is $\effgeo \approx 25\%$ according
to simulation studies. For \Lc to \Lz decays, \eg \LctoLpi, the geometrical efficiency is
 reduced by about a factor 50 since most \Lz baryons decay after the detector tracking volume.
%
%
 The \lhcb software-based trigger for the upgrade detector~\cite{LHCb-TDR-016} is expected to have efficiency for charm hadrons comparable
 to the current high level trigger~\cite{LHCb-DP-2014-002}, \ie $\efftrigger \approx 80\%$. A specific trigger scheme for the fixed-target
 experiment can be adopted to enhance the trigger efficiency for \Lc decays close to $100\%$.
 For example, a trigger based on the energy loss in a instrumented silicon crystal was used in the E761
 experiment to enhance the rate of reconstructed channeled \PSigmap baryons~\cite{Chen:1992wx}.
The tracking efficiency is estimated to be $70\%$ per track, leading to an efficiency $\efftrack \approx 34\%$ for
a \Lc decay with three charged particles.
 The detector reconstruction efficiency, $\effdet = \effgeo\efftrigger\efftrack$, is estimated to be 
\begin{align}
\effdet(\proton\Km\pip) &\approx 5.4\times10^{-2} \hspace{1cm} \text{ for } \Lctopkpi , \nonumber\\
\effdet(\Lz\pip) &\approx 1.0\times10^{-3} \hspace{1cm} \text{ for } \LctoLpi.
\end{align}
%
%
%

The initial \Lc polarization will be eventually measured using non-channeled \Lc particles.
Few \Lc decay asymmetry parameters are known, the only one relevant for our experiment is that associated to $\Lc\to\Lambda(p\pi^-)\pi^+$, $\alpha_{\Lambda\pip} = -0.91\pm 0.15$~\cite{Olive:2016xmw}.
Asymmetry parameters for different \Lc decays can be measured precisely at \lhcb in the future.
At present,
they can be computed from existing $\Lc\to\proton\Km\pip$ amplitude analysis results~\cite{Aitala:1999uq} (see Appendix~\ref{app:B}), 
yielding 
$\alpha_{\Deltares^{++}\Km} = -0.67 \pm 0.30$
for the \Lctodeltak decay. 
%


For the sensitivity studies we assume $s_0=0.6$ and $(g-2)/2 = 0.3$, according to experimental results and available theoretical predictions,
respectively, quoted in Ref.~\cite{Samsonov:1996ah}.
The $g-2$ and $d$ values can be derived from Eq.~\eqref{eq:EDM_LcPol} as
\begin{eqnarray}
&g-2& \approx \frac{2}{\gamma\theta_C}\arccos{\left(\frac{A_y}{\alpha s_0}\right)} 
      \approx \frac{2}{\gamma\theta_C}\arcsin{\left(\frac{A_z}{\alpha s_0}\right)} ,\\
&d & \approx \frac{(g-2)A_x}{\alpha s_0 \left[ \cos \Phi  -1 \right]},
\end{eqnarray}
where the quantity $A_{x,y,z}=\alpha s_{x,y,z}$ is measured from a fit to the angular
distribution of the decay products.
The main contribution to the statistical uncertainty on $g$ and $d$, in the limit $\gamma \gg 1$, can be estimated as
\begin{eqnarray}
&\sigma_{g}& \approx \frac{2}{\alpha s_0 \gamma \theta_C }\frac{1}{\sqrt{N_\Lc^{\rm reco}}},  \\
&\sigma_d& \approx \frac{g-2}{\alpha s_0\left[ \cos\Phi -1 \right]}\frac{1}{\sqrt{N_\Lc^{\rm reco}}},
\label{eq:EDM_stat_uncertainty}
\end{eqnarray}
where $N_\Lc^{\rm reco}$ is the number of channeled and reconstructed \Lc, as given in Eq.~(\ref{eq:NLcReco}), and $\Phi \approx 3 \rad$ 
is the precession angle defined in Eq.~(\ref{eq:MDM_angle})
estimated using the quantities reported in Table~\ref{tab:quantities}.
The estimate assumes negligibly small uncertainties on $\theta_C$, $\gamma$ and the initial \Lc polarization, $s_0$,
the latter to be measured with large samples of non-channeled \Lc decays.

Given the estimated quantities reported in Table~\ref{tab:quantities}, we obtain
\begin{align}
\frac{dN_\Lc^{\rm reco}}{dt} & \approx 5.9 \times 10^{-3}~\invs = 21.2~\invh \hspace{0.8cm} \text{for }\Lctodeltak, \nonumber\\
\frac{dN_\Lc^{\rm reco}}{dt} & \approx 8.3 \times 10^{-5}~\invs = 0.3~\invh \hspace{1cm} \text{for }\LctoLpi.
\end{align}
For reaching a sensitivity of $\sigma_d=0.01$, corresponding to a \Lc EDM of
$\delta = 2.1 \times 10^{-17} e \cm$,
we need, inverting Eq.~\eqref{eq:EDM_stat_uncertainty}, $5.6\times 10^3$ \Lctodeltak or $3.0\times 10^3$ \LctoLpi events, recorded during a data taking time $t$ of
\begin{align}
t &= 265 {~\rm h} = 11 {~\rm days}  \hspace{4.35cm} \text{for }\Lctodeltak, \nonumber\\
t &= 1.0\times 10^{4}{~\rm h} \approx 420 {~\rm days} \approx 1.2 {~\rm years}  \hspace{1cm} \text{for }\LctoLpi.
\end{align}
%
Therefore, a
measurement of \Lc EDM is feasible 
in \Lc quasi two-body decays at \lhcb, while it is difficult in \Lc to \Lz final states.

\begin{figure}
\centering
\includegraphics[scale=0.36]{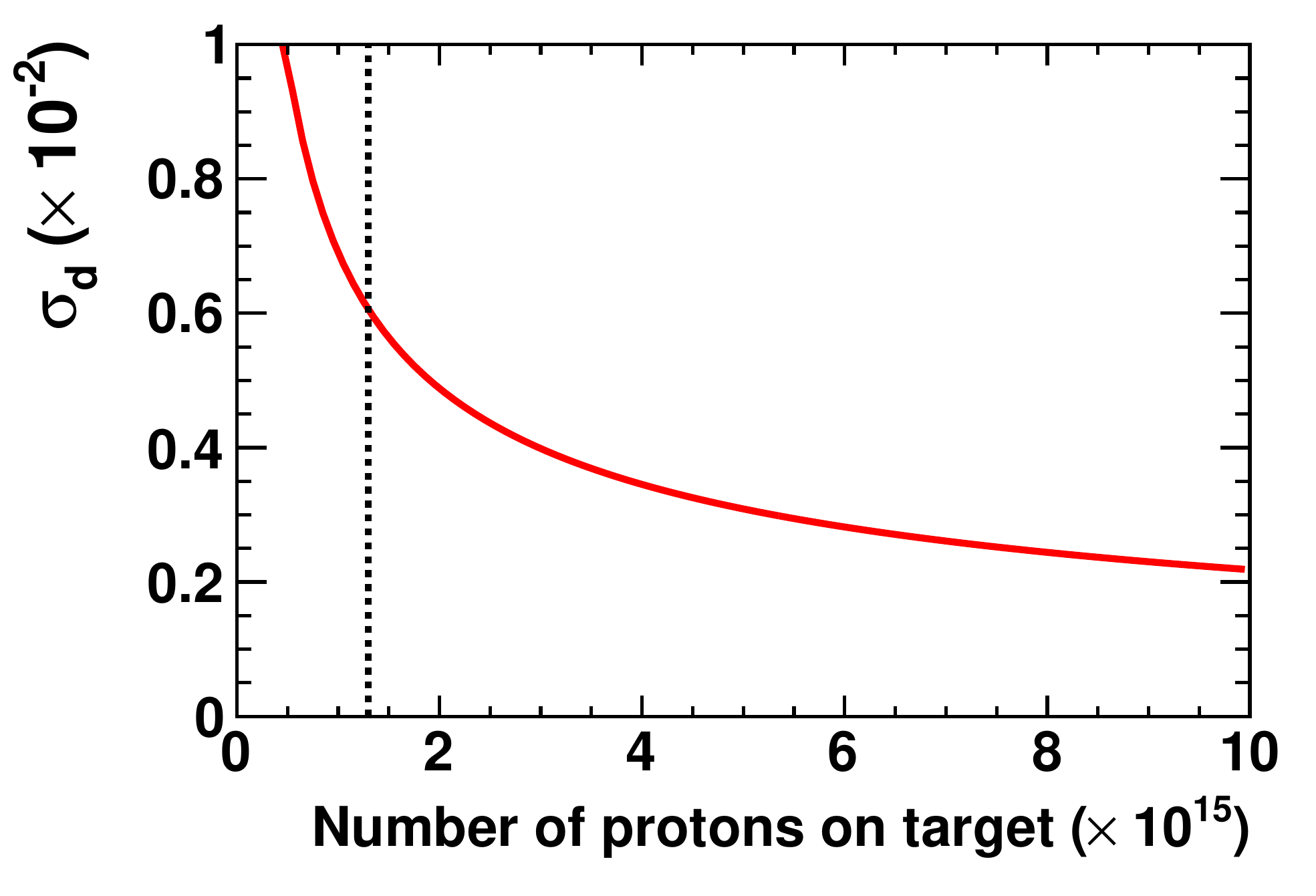}
\includegraphics[scale=0.36]{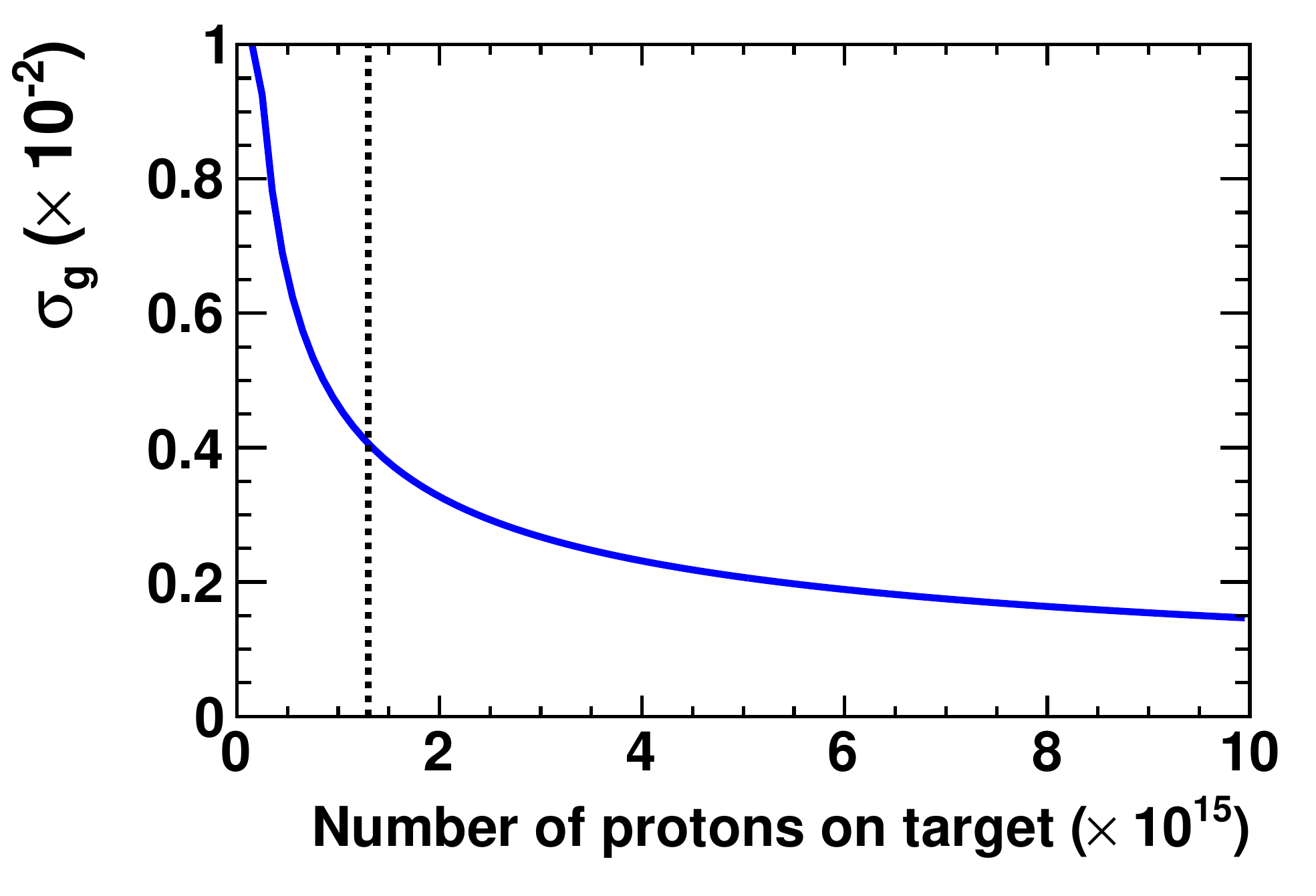}
\caption{Dependence of the (Left) $d$ and (Right) $g$ uncertainties for the
  \Lc baryon, reconstructed in $\Deltares^{++}\Km$ final state, with the number of protons on target.
  One month of data taking corresponds to $1.3\times 10^{15}$ incident protons (dashed line),
   according to the estimated quantities listed in Table~\ref{tab:quantities}. 
\label{fig:Lambdac_sensitivity}}
\end{figure}

Considering only \Lctodeltak events, the uncertainties scale as
\begin{eqnarray}
 \sigma_g \approx 4.0\times 10^{-3}\frac{1}{\sqrt{t({\rm month})}}~,~~ 
 \sigma_d \approx 6.1\times 10^{-3}\frac{1}{\sqrt{t({\rm month})}},
\end{eqnarray}
corresponding to
\begin{eqnarray}
 \sigma_{\mu} \approx 4.2 \times 10^{-27} \textrm{erg/G} \frac{1}{\sqrt{t({\rm month})}}~,~~  
 \sigma_{\delta} \approx 1.3 \times 10^{-17} e \cm \frac{1}{\sqrt{t({\rm month})}},
\end{eqnarray}
where the time $t$ of the data taking period is expressed in months. 
The dependence of the sensitivity to \Lc EDM and MDM as a function of the number of incident protons on the target is shown in Fig.~\ref{fig:Lambdac_sensitivity}.

Estimating the \Xicp baryon production and the absolute $\Xicp\to\proton\Km\pip$
branching fraction as described in Sec.~\ref{sec:method_lambda}, we obtain the ratio 
\begin{equation}
\frac{\sigma_\Xicp\br(\Xicp\to pK^-\pi^+)}{\sigma_\Lc\br(\Lc\to pK^-\pi^+)} \approx 18 \%,
\end{equation}
while the fraction of \Xicp baryons decaying after the crystal is $\effDF(\Xicp) \approx 47\%$. Assuming decay asymmetry parameters and initial polarization similar to the \Lc baryon, the expected statistical uncertainty on the \Xicp MDM and EDM is
\begin{eqnarray}
%
\sigma_{\mu} \approx 6.3 \times 10^{-27} \textrm{erg/G} \frac{1}{\sqrt{t({\rm month})}}~,~~
\sigma_{\delta} \approx 2.0 \times 10^{-17} e \cm \frac{1}{\sqrt{t({\rm month})}}.  
\end{eqnarray}

\section{Conclusions}
\label{sec:conclusion}
The unique possibility to search for the EDM of strange and charm baryons at LHC is discussed,
based on the  exploitation of large statistics of baryons with large Lorentz boost and polarization.
The \Lz strange baryons are selected from weak charm baryon decays produced in \proton\proton collisions at
$\approx 14$~\tev center-of-mass energy, while \Lc and \Xicp charm baryons are produced in a fixed-target experiment to be installed in the LHC,
in front ot the \lhcb detector. Signal events can be reconstructed using the \lhcb detector in
both cases.
The sensitivity to the EDM and  the MDM of the strange and charm baryons arises from the study of the spin precession
in intense electromagnetic fields.
The long-lived \Lz precesses in the magnetic field of the detector tracking system.
Short-lived charm baryons are channeled in a bent crystal attached to the target and the intense electric field
between atomic planes induces the spin precession.   
Sensitivities for the \Lz EDM at the level of $ 1.3 \times 10^{-18}~e\cm$ 
can be achieved
using a data sample corresponding to an integrated luminosity of 50 \invfb to be collected
during the LHC Run 3. A test of \CPT symmetry can be performed by measuring
the MDM of \Lz and \Lbar baryons with a precision of about $4\times 10^{-4}$ on the $g$ factor.
The EDM of the \Lc (\Xicp) can be searched for with a sensitivity of $1.3~(2.0)\times 10^{-17}/\sqrt{t({\rm month})}~e\cm$ with 
dedicated runs or running in synergetic mode with the \lhcb experiment, in parallel to \proton\proton collisions.
Both solutions have to be studied in details using ad-hoc simulations.
The proposed experiment would allow about two orders of magnitude improvement in the sensitivity for the
\Lz EDM and the first search  for the charm baryon EDM, 
expanding
the search for new physics
through the EDM of fundamental particles.

\section*{Acknowledgements}
 
\noindent We express our gratitude to our colleagues of the \lhcb collaboration.
The authors would like to thank G.~Cavoto, M.~Ferro-Luzzi, G.~Graziani, M.~Nebot, M.~Schiller,
A.~Pich, V.~Vagnoni and G.~Wilkinson for interesting discussions.
We acknowledge support from INFN (Italy), MinECo and GVA (Spain).

%

%
\appendix
\section{Spin precession and time evolution equations}
\label{app:A}
The time evolution of the spin-polarization vector for a particle with charge $q$
in an electromagnetic field, as a function of the proper time $\tau$,   
is given by the Thomas-Bargmann-Michel-Telegdi 
(T-BMT) equation~\cite{Thomas:1926dy,Thomas:1927yu,Bargmann:1959gz},
\begin{equation}
\label{eq:TBMTCov}
\frac{da^\mu}{d\tau} 
= \frac{g \mu_B}{\hbar} \left[ F^{\mu\nu}a_\nu + \frac{1}{c^2} \left( a_\alpha F^{\alpha\beta} u_\beta \right) u^\mu \right]
- \frac{1}{c^2} \left( a_\alpha \dot{u}^\alpha \right) u^\mu 
- \frac{d \mu_B}{\hbar} \left[  F^{*\mu\nu}a_\nu + \frac{1}{c^2} \left( a_\alpha F^{*\alpha\beta} u_\beta \right) u^\mu  \right] ,
\end{equation}
where $F^{\mu\nu}$ is the electromagnetic tensor, $a^\mu = (a^0,\mathbf a)$ is the spin
4-pseudovector, and $p^\mu = m u ^\mu = \left(E/c,\mathbf p \right)$ is
the momentum 4-vector.
For homogeneous fields,
the  velocity derivative is given by the Lorentz force,
\begin{equation}
\dot{u}^\mu \equiv \frac{du^\mu}{d\tau} = \frac{q}{mc} F^{\mu\nu} u_\nu .
\end{equation}
In the rest frame of the particle, $a^\mu=(0, \mathbf s)$, $p^\mu=(mc,\bm 0)$, where $\mathbf s$ is the non-relativistic spin-polarization 
vector.
Therefore, in any frame $a^\mu p_\mu=0$ and $a_\mu a^\mu=-{\mathbf s}^2$.

In a frame comoving with respect to the particle rest frame where the particle has velocity $\bm \beta = \mathbf p / m \gamma $,
\eg the laboratory frame,
$a^\mu$ is given by~\cite{Jackson:1998nia,Leader2011}
\begin{equation}
\mathbf a = \mathbf s + \frac{\gamma^2}{\gamma+1} (\bm \beta \cdot \mathbf s) \bm \beta~,~~
 a^0 = \bm \beta \cdot \mathbf a = \gamma(\bm \beta \cdot \mathbf s) ,
\label{eq:SpinLab}
\end{equation}
where the components of the momentum 4-vector are $p^0 = \gamma m c^2$ and $\mathbf p =\gamma m \bm \beta c$. 
Substituting in the covariant Eq.(\ref{eq:TBMTCov}), the
spin precession equation is~\cite{Jackson:1998nia,Leader2011,Fukuyama:2013ioa,Silenko:2014uca},
\begin{equation} 
\label{eq:TBMTgeneral}
\frac{d \mathbf s}{ d t} =  \mathbf s \times \bm \Omega ~, ~~~ \bm \Omega = \bm \Omega_{\rm MDM} + \bm \Omega_{\rm EDM} + \bm \Omega_{\rm TH} ,
\end{equation}
where $t$ is the time in the laboratory frame, and
the precession angular velocity vector $\bm \Omega$ has been split into three contributions,
\begin{equation} 
\label{eq:OMEGAgeneral}
\bm \Omega_{\rm MDM} 
 = \frac{g \mu_B}{\hbar} \left( \mathbf B - \frac{\gamma}{\gamma+1}(\bm \beta \cdot \mathbf B)\bm \beta - \bm \beta \times \mathbf E\right) , 
\end{equation}
\begin{equation}
\nonumber
\bm \Omega_{\rm EDM} 
= \frac{d \mu_B}{\hbar} \left( \mathbf E - \frac{\gamma}{\gamma+1}(\bm \beta \cdot \mathbf E)\bm \beta + \bm \beta \times \mathbf B\right) , 
\end{equation}
\begin{equation}
\nonumber
 \bm \Omega_{\rm TH}
= \frac{\gamma^2}{\gamma+1} \bm \beta  \times \frac{d \bm \beta}{d t}
= \frac{q}{mc} \left[ \left( \frac{1}{\gamma} - 1 \right) \mathbf B + \frac{\gamma}{\gamma + 1} (\bm \beta \cdot \mathbf B)\bm \beta  -  \left(  \frac{1}{\gamma+1} -1 \right) \bm \beta \times \mathbf E  \right],
\end{equation}
corresponding to the MDM, EDM and Thomas precession.
The electric and magnetic fields, $\mathbf E$ and $\mathbf B$, respectively, are expressed in the laboratory frame.

For a neutral particle ($q=0$) the Thomas precession term, arising from Lorentz forces, does not contribute and we obtain the classical 
equation, $d \mathbf s/ d \tau = \bm \mu \times \mathbf B^* + \bm \delta \times \mathbf E^*$,
where $\mathbf E^*$ and $\mathbf B^*$ are the external fields in the rest frame of the particle~\cite{Jackson:1998nia}. 
%
Equations~(\ref{eq:TBMTgeneral}) and~(\ref{eq:OMEGAgeneral}) can be generalized to account for field gradient effects as described in Ref.~\cite{Good:1962zza,Metodiev:2015gda}. 
%

\subsection{Spin time evolution for the \Lz case}
\label{app:lambda}

For $\mathbf E = 0$ and $q=0$, Eqs.~(\ref{eq:TBMTgeneral}) and (\ref{eq:OMEGAgeneral}) simplify to 
\begin{equation}\label{eq:TBMTLHCb}
\frac{d \mathbf s}{ d t} = \mathbf s \times \bm \Omega , 
\end{equation}
\begin{equation}\label{eq:TBMTLHCb-Omega}
%
\bm \Omega
= \frac{\mu_B}{\hbar} \left[ g \left( \mathbf B -\frac{\gamma}{\gamma+1}(\bm \beta \cdot \mathbf B)\bm \beta \right)  + d \bm \beta \times \mathbf B  \right] ,
\end{equation}
where $\bm \beta$ is the particle velocity in the laboratory frame.
%
This system of homogeneous first order linear differential equations
can be solved analytically with the approximation that the precession of the particle depends only on 
the integrated magnetic field along its flight path.
Given the initial condition $\mathbf s(0)=\mathbf s_0$, the time evolution of the polarization is
\begin{equation}\label{eq:sAnalytical}
\mathbf s (t)
 = (\mathbf s_0  \cdot \bm \omega) \bm \omega 
 + \left[ \mathbf s_0 - (\mathbf s_0 \cdot \bm \omega) \bm \omega \right] \cos(\Omega t) 
 + (\mathbf s_0 \times \bm \omega) \sin (\Omega t) , 
\end{equation}
where $\Omega = |\bm \Omega|$ and $\bm \omega = \bm \Omega / \Omega$, with the
precession angular velocity given by Eq.~(\ref{eq:TBMTLHCb-Omega}).

The polarization in terms of the experimentally measured \Lz flight length $l=\beta c t$, $\mathbf s (l)$, has similar form,
\begin{equation}\label{eq:sAnalytical-DL}
\mathbf s (l)
 = (\mathbf s_0  \cdot \bm \omega') \bm \omega' 
 + \left[ \mathbf s_0 - (\mathbf s_0 \cdot \bm \omega') \bm \omega' \right] \cos\Phi 
 + (\mathbf s_0 \times \bm \omega') \sin\Phi , 
\end{equation}
where $\Phi = |\bm \Phi|$ and $\bm \omega' =\bm \Phi/\Phi$. 
The precession angle vector is	
\begin{equation}\label{eq:sAnalytical2}
   \bm \Phi
	= \frac{\mu_B}{\beta \hbar c} \left[ g \left( \mathbf D -\frac{\gamma}{\gamma+1}(\bm \beta \cdot \mathbf D)\mathbf \beta \right) + 
                 d \bm \beta \times \mathbf D \right] ,
\end{equation}
with $\mathbf D \approx \mathbf{\overline B} l = \int_0^l \mathbf B(\mathbf r_0 + \bm \beta l'/\beta)dl'$ the integrated magnetic field 
along the \Lz flight path.
%
%
%
\subsubsection{Magnetic field gradients}
\label{app:lambda_BfieldGrad}
%
%
The inhomogeneities of the magnetic field are not expected to introduce significant effects in the spin precession. 
The spin equation of motion including first-order field gradients is derived in Ref.~\cite{Metodiev:2015gda} to be
%
\begin{align}
\bm{\Omega}_{\rm MDM} &= \frac{g\mu_B}{\hbar} \left[ \bf{B} - \frac{\gamma}{\gamma+1} (\bm{\beta}\cdot\bf{B}) \bm{\beta}\right] \nonumber\\
&+ \frac{g\mu_B}{2}\frac{1}{mc} \frac{\gamma}{\gamma+1} (\bm{\beta}\times\bm{\nabla}) \left[\bm{s}\cdot\left(\bf{B}-\frac{\gamma}{\gamma+1} \bm{\beta}(\bm{\beta}\cdot\bf{B})\right) \right]~, \nonumber\\
\bm{\Omega}_{\rm EDM} &= \frac{d\mu_B}{\hbar}  \left[ \bm{\beta}\times\bf{B} \right] +
\frac{d\mu_B}{2}\frac{1}{mc} \frac{\gamma}{\gamma+1} (\bm{\beta}\times\bm{\nabla})\left[\bm{s}\cdot(\bm{\beta}\times\bf{B})\right] .
\end{align}
In \lhcb the ratio of the field gradient terms to the homogeneous field ones can be estimated
as 
$$
\frac{\hbar}{2 m c}\frac{\beta \gamma}{\gamma+1}  \frac{|\bm{\nabla}B|}{B}\sim 7.4 \times 10^{-16}~,$$
with $\beta\simeq 1$ and $\gamma \gg 1$, and where $|\bm{\nabla}B| = 1.14~\text{Tm}^{-1}$ and
$B=1~\text{T}$ are the maximum values within the detector acceptance as extracted from the
\lhcb field mapping~\cite{LHCb-DP-2014-002,Hicheur:2007jfk}. 
Therefore, this effect is negligibly small at \lhcb.


\subsubsection{Spin rotations}
\label{app:lambda_rotations}

\begin{figure}[h!]
\centering
{ \includegraphics[width=0.85\linewidth]{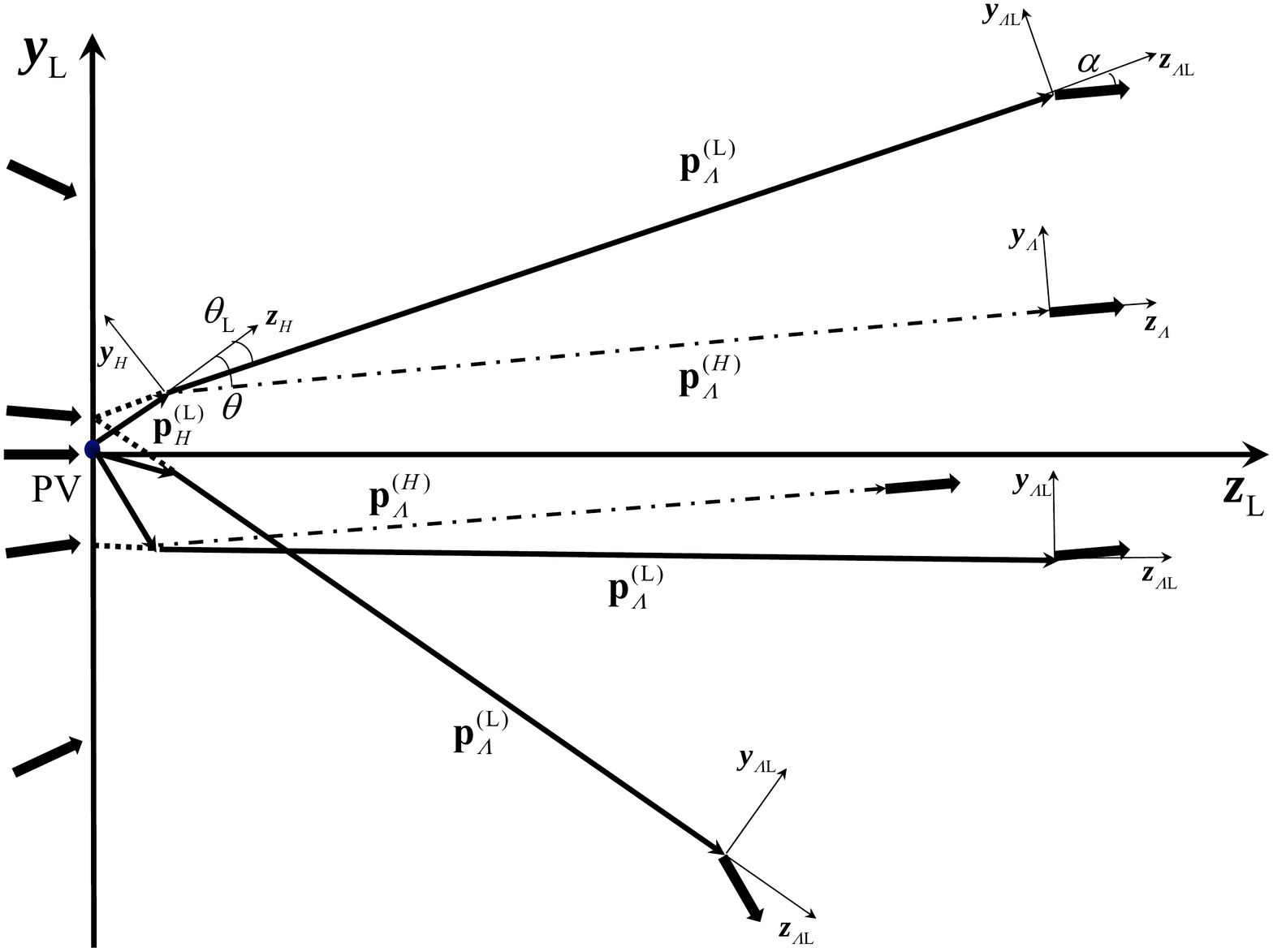} }
\caption{Sketch of the heavy baryon production at the primary vertex (PV) and its decay into a \Lz, showing the \SH, \SLz and \SLzL
helicity frames, in the $zy$ plane in \SL. Continuous (dotted-dashed) arrows represent momenta in \SL (\SH) frame.
The \Lz polarization vector (thick arrow at the right) is aligned along the $z$ axis in \SLz (longitudinal polarization), 
and rotated by the Wick angle $\alpha$ with respect to $z$ in \SLzL.
The polarization state of the \Lz in \SLzL (thick arrows at the left) is correlated with its apparent production point on the 
$z$ plane in \SL intersecting the PV.
These points are shown by the short-dashed lines traced back from the \Lz trajectory (intersecting the \PH decay point). The angle
$\theta$ ($\theta_{\mathrm L}$) is formed by the \Lz momentum in the \SH (\SL) frame
with respect to the $z$ axis in \SH.
}
\label{fig:frames-rotations}
\end{figure}

The variation of the \Lz momentum direction in the laboratory frame results in an initial polarization vector which is not 
fixed to be perpendicular to the magnetic field. The relative orientation of the spin and magnetic field vectors is determined 
by two rotations. 
On one hand, the polarization vector from the equation of motion is given in the comoving rest frame
reached from the laboratory frame, \SL,
by a pure boost. This is usually referred to as canonical frame~\cite{Leader2011}.
However, the analyser, given by Eq.~\ref{eq:AngDist}, is defined in the particle helicity frame. 
The two rest frames, canonical and helicity, are related by the rotation between the \SL and \SLzL frames, 
defined by the \Lz and \PH momentum directions in \SL (see Fig.~\ref{fig:frames}).
One the other hand, the choice of the \SLzL frame induces a second rotation of the polarization components with respect
to the \SLz frame, where the \Lz longitudinal polarization is maximal, as illustrated in Fig.~\ref{fig:frames-rotations}.
This is known in the literature as Wick rotation.
To avoid dilution effects,
the change of the polarization 
has to be analysed as a function of the
kinematics of the decay.
%
For example, a longitudinally polarized \Lz with polarization $s_0$ along $z$ in \SLz would 
have a transverse component in \SLzL of magnitude $s_0 \sin\alpha$,
with $\sin\alpha = (m_\Lz / m_{\PH}) (p_{\PH}^{(\mathrm L)} / p_\Lz^{\mathrm (L)}) \sin \theta$~\cite{Leader2011}.
As shown in Fig.~\ref{fig:frames-rotations}, 
the \Lz helicity angle $\theta$ and the spin direction are related to the \Lz impact parameter in 
the laboratory~\cite{Grosnick:1989qv}.
The relation can be exploited to define ensembles of \Lz particles having similar initial polarization, 
therefore improving the sensitivity to detect the spin change.

For the sensitivity studies, the rotation of the magnetic field into the \SLzL frame and
the Wick rotation are neglected. The first is expected to have a negligible
impact on our study since \Lz baryons have momenta
largely along the $z$ axis, and the main component of the magnetic field is 
along the vertical direction ($B_y$), thus mostly perpendicular to the \Lz motion.
Instead, the effect of the Wick rotation is not relevant when measuring the spin change of 
ensembles of \Lz particles having similar initial polarization.


\subsection{Spin time evolution for the \Lc and \Xicp case}
\label{app:lambdac}
%
For $\mathbf B = 0$ and $q = +1$, \equationname~\eqref{eq:OMEGAgeneral} simplifies to
\begin{equation}
\mathbf \Omega = \frac{2\mu'}{\hbar} \left( \mathbf E\times\bm\beta \right) + \frac{d\mu_B}{\hbar} \mathbf E +\frac{1}{\gamma+1} \frac{2\mu_B}{\hbar} \left( \mathbf E\times\bm \beta \right) - \frac{d\mu_B}{\hbar} \frac{\gamma}{\gamma+1} \left( \bm\beta\cdot\mathbf E \right) \bm\beta, \label{eq:TBMT_channelled}
\end{equation}
where 
\begin{equation}
\mu' = \frac{g-2}{2}\frac{e\hbar}{2m c},
\label{eq:mu_prime}
\end{equation}
is the anomalous magnetic moment for a spin-1/2 particle.
Since we are dealing with ultra relativistic \Lc with $\gamma \approx 437$ at $1 \tev$ energy,
in first approximation the terms $\propto 1/\gamma$ are neglected.
\begin{figure}
\centering
\includegraphics[scale=0.7]{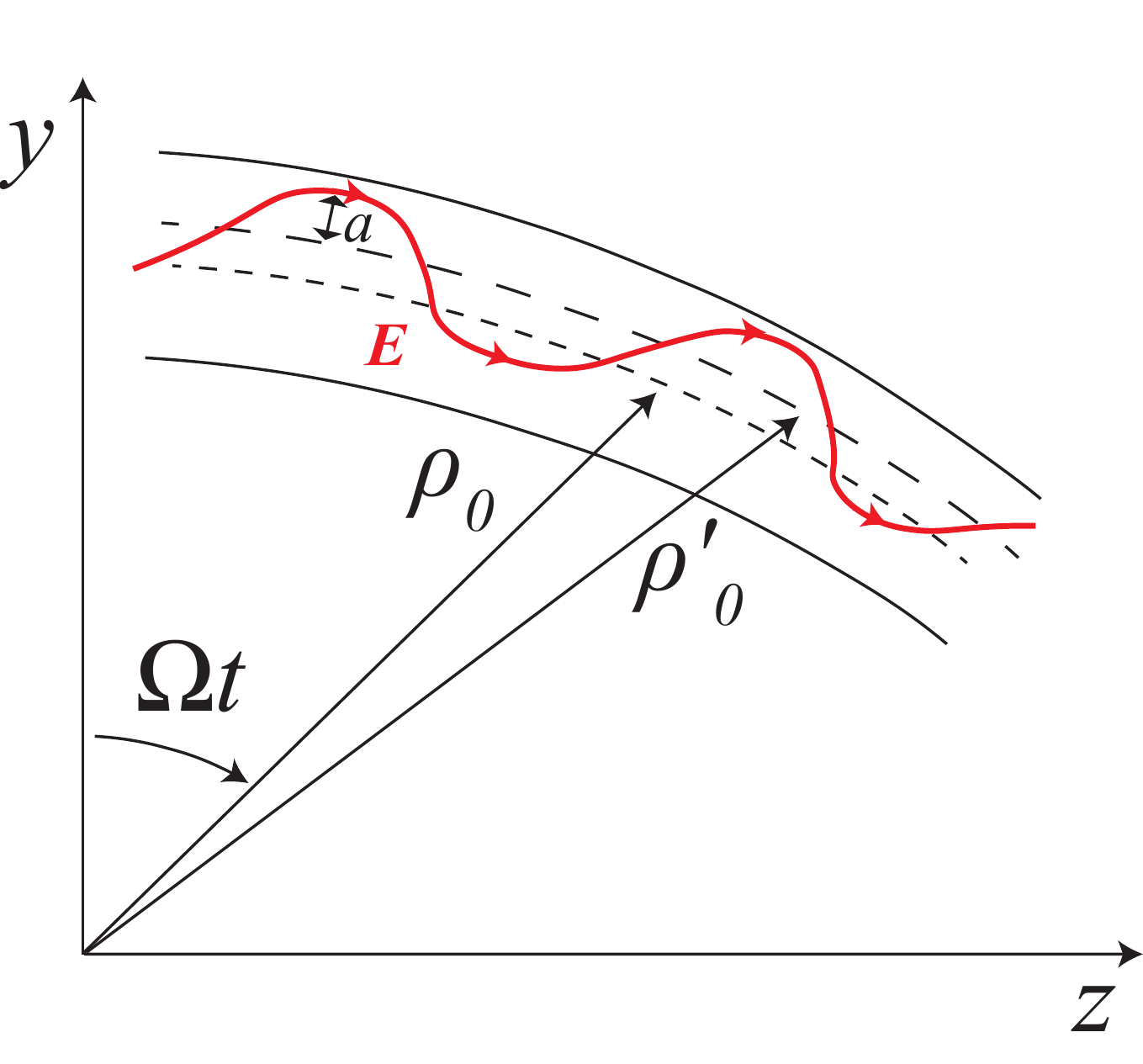}
\caption{Radial coordinates definition:
  $\rho_0$ is the radius corresponding to the minimum of the harmonic electric potential; $\rho_0'$
  represents the radial equilibrium position of the electric and centrifugal potential.
  The red curve represents the particle trajectory inside the crystal in
  presence of the radial electric field $\mathbf{E}$, $a$ is the oscillation amplitude and
  $\Omega$ the revolution frequency.
  \label{fig: trajectory in crystals}}
\end{figure}

We  describe the particle trajectory in a bent crystal using radial coordinates~\cite{Baryshevsky:2015zba}, as shown in Fig.~\ref{fig: trajectory in crystals},
\begin{equation}
x(t) = {~\rm const.}, \hspace{1cm} y(t) = \rho(t)\cos(\Omega t), \hspace{1cm} z(t) = \rho(t)\sin(\Omega t),
\end{equation}
where $\Omega$ is the revolution frequency for the particle traversing the bent crystal.
In our ultra-relativistic case it is well approximated by $\Omega\approx c/\rho_0$,
where $\rho_0$ is the crystal curvature radius. 
The radius of the trajectory as a function of time is 
\begin{equation}
\rho(t) = \rho'_0 + a \cos(\Omega_k t + \delta),
\label{eq:rho(t)}
\end{equation}
where $a$, $\Omega_k$ and $\delta$ are the oscillation amplitude, frequency and phase, respectively; $a$ and $\delta$ depend on the particle energy and incident angle,
while $\Omega_k$ depends on the crystal potential and particle energy.
The radial equilibrium position $\rho'_0$ differs from the electric potential minimum
position $\rho_0$, due to the centrifugal potential, avoiding periodical cancellations and
 therefore inducing spin precession~\cite{Kim:1982ry}.
The electric potential in the crystal around the minimum can be approximated as an harmonic
  potential,
\begin{equation}
V = \frac{k}{e} \frac{\left[ \rho(t) - \rho_0 \right]^2}{2},
\label{eq:harmonic_potential}
\end{equation}
and the corresponding electric field is 
\begin{equation}
E_x = 0 \hspace{1cm} E_y=-\frac{dV}{d\rho} \cos(\Omega t) \hspace{1cm} E_z = -\frac{dV}{d\rho} \sin(\Omega t),
\end{equation}
where the oscillation frequency of the particle around its equilibrium position $\rho'_0$ is
$\Omega_k=\sqrt{kc^2/eW}$ with $W$ being the particle energy. Typical values for the
relevant quantities are $\rho_0 \sim 30 \m$, $\Omega\approx c/\rho_0 \sim 10^7 \hz$, $a \sim 10^{-10} \m$, $k = 4\times 10^{17} \ev/\cma$ for a Si crystal,
yielding $\Omega_k \sim 10^{13} \hz$ for $1 \tev$ particles.


Substituting the radial coordinates and applying the ultra-relativistic approximation to Eq.~(\ref{eq:TBMT_channelled}) we obtain:
\begin{align}
\Omega_x &\approx \frac{2\mu'}{\hbar} (E_y\beta_z - E_z\beta_y) = \frac{2\mu'}{\hbar} \left(- \frac{dV}{d\rho} \frac{\rho\Omega}{c} \right) \nonumber\\
\Omega_y &\approx \frac{d\mu_B}{\hbar} \left[ E_y - \left( \bm\beta\cdot\mathbf E \right) \beta_y \right] = \frac{d\mu_B}{\hbar} \left\{ -\frac{dV}{d\rho}\cos(\Omega t) + \frac{dV}{d\rho} \frac{\dot{\rho}}{c^2} \left[ -\rho\Omega\sin(\Omega t) + \dot{\rho}\cos(\Omega t) \right] \right\} \nonumber\\
\Omega_z &\approx \frac{d\mu_B}{\hbar} \left[ E_z - \left( \bm\beta\cdot\mathbf E \right) \beta_z \right] = \frac{d\mu_B}{\hbar} \left\{ -\frac{dV}{d\rho}\sin(\Omega t) + \frac{dV}{d\rho} \frac{\dot{\rho}}{c^2} \left[ \rho\Omega\cos(\Omega t) + \dot{\rho}\sin(\Omega t) \right] \right\}.\label{eq:precession_vector}
\end{align}

In absence of EDM, \ie $d=0$, the spin precession inside the bent crystal
occurs in the $yz$ plane with the following spin time evolution~\cite{Baryshevsky:2015zba},
\begin{equation}
\mathbf s(t) ~=~
\left\lbrace
\begin{array}{l}
s_x(t) = 0 \\
s_y(t) = s_0 \cos\left(\omega t\right) \\
s_z(t) = - s_0 \sin\left( \omega t\right)
\end{array}
\right. ,
\label{eq:spin_precession_zero_EDM}
\end{equation}
for the initial condition ${\mathbf s_0} = \left( 0, s_0, 0\right)$ and
where $\omega \approx 2\mu'E(\rho'_0)/\hbar $ is the precession frequency.
The spin precession angle defined in \equationname~\eqref{eq:MDM_angle} is
$\Phi= \omega\overline{t}$, where $\overline{t}$ is the time needed to traverse the crystal.  
In presence of a non-zero EDM the spin precession is no longer confined to the $yz$ plane,
generating a $s_x$ spin component otherwise not present,
%
%
\begin{eqnarray}
\frac{ds_x}{dt} & = & s_y\Omega_z - s_z\Omega_y \nonumber\\
& \approx  & \frac{d\mu_B}{\hbar} \frac{dV}{d\rho} s_0 \bigg\{-\cos(\omega t)\sin(\Omega t) - \sin(\omega t)\cos(\Omega t) \nonumber\\
&  & \,\,\,\,\,\,\,\,\,\,\,\,\,\,\,\,\,\,\,\,\,\,\,\,\,\,\,\,\,\,\,\,
     +\frac{\dot{\rho}\rho\Omega}{c^2} \big[ \cos(\omega t)\cos(\Omega t) - \sin(\omega t)\sin(\Omega t) \big] \nonumber\\
&  & \,\,\,\,\,\,\,\,\,\,\,\,\,\,\,\,\,\,\,\,\,\,\,\,\,\,\,\,\,\,\,\,
     + \left.\frac{\dot{\rho}^2}{c^2}\big[ \cos(\omega t)\sin(\Omega t) + \sin(\omega t)\cos(\Omega t) \big] \right\rbrace \nonumber\\
& = & \frac{d\mu_B}{\hbar} \frac{dV}{d\rho} s_0 \bigg\{ -\sin\left[(\omega+\Omega)t\right] + \frac{\dot{\rho}\rho\Omega}{c^2} \cos\left[(\omega+\Omega)t\right] + \frac{\dot{\rho}^2}{c^2} \sin\left[(\omega+\Omega)t\right] \bigg\}.
\label{eq:sy_evolution}
\end{eqnarray}
To derive Eq.~(\ref{eq:sy_evolution}), EDM effects are assumed to be small compared to the
MDM effects, \ie $d \ll (g-2)$, and therefore $\Omega_y,\Omega_z\ll\Omega_x$.
We neglect terms of order $\dot{\rho}/c$ where
\begin{equation}
\dot{\rho} = -a\Omega_k \sin(\Omega_k t + \delta) \sim a\Omega_k \sim 10^3 \m/\sec,
\end{equation}
since the second term of \equationname~\eqref{eq:sy_evolution} is about $ \dot{\rho}\rho\Omega/c^2 \sim \dot{\rho}/c \sim 3 \times 10^{-4}$ and the third term is about $\dot{\rho}^2/c^2 \sim 9 \times 10^{-8}$. 
We demonstrate that $\Omega \ll \omega$ by requiring the electric force to be identical to
the centripetal force, 
\begin{equation}
\frac{m\gamma c^2}{\rho'_0} = eE(\rho'_0),
\end{equation}
and obtain $\omega \approx \frac{2\mu'}{\hbar}E(\rho'_0) \sim 10^{10} \hz \gg \Omega \sim 10^7 \hz$.

Then, \equationname~\eqref{eq:sy_evolution} simplifies as
\begin{equation}
\frac{ds_x}{dt} = \frac{d\mu_B}{\hbar} \left(-\frac{dV}{d\rho}\right) s_0 \sin(\omega t),
\label{eq:sy evolution approximated}
\end{equation}
%
and the time evolution is
%
\begin{equation}
s_x(t) = -\frac{d\mu_B}{\hbar} E(\rho'_0) \int^t_0 \sin(\omega t') dt' - \frac{d\mu_B}{\hbar} \frac{ka}{e} \int^t_0 \cos(\Omega_k t'+\delta)\sin(\omega t') dt'.
\end{equation}
%
The second integral is negligibly small since $\Omega_k \gg \omega$ and its fast oscillation
averages the integral to zero. The calculation can be decomposed into two analytically
integrable terms proportional to $\sin(\Omega_k t')\sin(\omega t')$ and $\cos(\Omega_k t')\sin(\omega t')$. Assuming $\Omega_k \gg \omega$, the maximum value of this integral is
\begin{equation}
\sim \frac{d\mu_B}{\hbar}\frac{ka}{e\Omega_k} \sim 2\frac{d}{g-2}\xi,
\end{equation}
where $\xi = \mu'ka/\hbar e\Omega_k\lesssim 10^{-2}$ and terms proportional to $\xi$ were neglected to 
derive Eq.~\eqref{eq:spin_precession_zero_EDM}~\cite{Baryshevsky:2015zba}.
Finally we obtain the time evolution of the polarization vector in presence of a
non-negligible EDM,
\begin{equation}
\mathbf s(t) ~=~
\left\lbrace
\begin{array}{l}
s_x(t) \approx s_0 \dfrac{d}{g-2} \big[ \cos(\omega t) - 1 \big] \\
s_y(t) \approx s_0 \cos\left(\omega t\right) \\
s_z(t) \approx - s_0 \sin\left( \omega t\right)
\end{array}
\right. .
\end{equation}
%

\subsubsection{Electric field gradients}

The equations describing the particle trajectory and its spin precession in an electromagnetic field, including first-order electromagnetic field gradients, as well as a particle EDM contributions, are derived in \cite{Metodiev:2015gda}. In absence of magnetic fields the spin precession 
vector $\mathbf\Omega = \mathbf\Omega_{\rm MDM} + \mathbf\Omega_{\rm EDM} + \mathbf\Omega_{\rm TH}$ is
\begin{align}
\mathbf\Omega_{\rm MDM} &= \frac{g\mu_B}{\hbar} \left[ \mathbf E\times \bm\beta \right] + \frac{g\mu_B}{2}\frac{1}{mc} \frac{\gamma}{\gamma+1} (\bm\beta\times\mathbf\nabla)\left[\mathbf s\cdot(\mathbf E\times\bm\beta)\right], \nonumber\\
\mathbf\Omega_{\rm EDM} &= \frac{d\mu_B}{\hbar} \left[ \mathbf E - \frac{\gamma}{\gamma+1} (\bm\beta\cdot\mathbf E) \bm\beta\right] \nonumber\\
&+ \frac{d\mu_B}{2}\frac{1}{mc} \frac{\gamma}{\gamma+1} (\bm\beta\times\mathbf\nabla) \left[\mathbf s\cdot\left(\mathbf E - \frac{\gamma}{\gamma+1} \bm\beta(\bm\beta\cdot\mathbf E)\right) \right],
\end{align}
with unchanged Thomas precession component. Using the harmonic potential approximation we
obtain
\begin{equation}
\frac{d|\mathbf E|}{d\rho} = \frac{k}{e},
\end{equation}
and employing the values used in this appendix, the ratio of the field gradient terms
to the homogeneous field ones is estimated to be
\begin{equation}
\frac{\hbar\frac{d|\mathbf E|}{d\rho}}{2mc|\mathbf E|} = \frac{\hbar k \rho'_0}{2m^2\gamma c^3} \sim 2.3 \times 10^{-3} \, \frac{1}{\gamma},
\end{equation}
which is negligibly small in the ultra-relativistic regime.

When including electric field gradient effects, in absence of magnetic fields, the particle trajectory equation becomes
\begin{align}
mc\frac{d(\gamma\bm\beta)}{dt} &= q\mathbf E \nonumber\\
&+ \gamma^2\frac{g\mu_B}{2} \left[ \mathbf\nabla + \bm\beta\times(\bm\beta\times\mathbf\nabla) + \frac{1}{c}\bm\beta\frac{\partial}{\partial t}\right]\left[\mathbf s\cdot(\mathbf E\times\bm\beta)\right] \nonumber\\
&+ \gamma^2 \frac{d\mu_B}{2} \left[ \mathbf\nabla + \bm\beta\times(\bm\beta\times\mathbf\nabla) + \frac{1}{c}\bm\beta\frac{\partial}{\partial t}\right] \left[\mathbf s\cdot\left(\mathbf E-\frac{\gamma}{\gamma+1} \bm\beta(\bm\beta\cdot\mathbf E)\right) \right],
\end{align}
where the first term is the Lorentz force and the following two terms are the MDM and EDM contributions. In our experimental setup the initial spin vector is orthogonal to $\mathbf E\times\bm\beta$,
hence the MDM component is negligible. The typical magnitude of the ratio between the EDM electric field gradient term and the Lorentz force contribution is $\sim d\gamma \times 10^{-3}$ which can be close to 1 for $\gamma\sim 1000$ only if $d\sim 1$, \ie similar EDM and MDM magnitudes. However, we assume the EDM magnitude to be tiny with respect to the MDM one, as already assumed
in the derivation of the spin equation of motion. In case of a large EDM, this term would
make the spin precession frequency dependent on the spin direction.

\section{Asymmetry parameter $\alpha$ for quasi two-body final states
in \decay{\Lc}{\proton\Km\pip} decays}
\label{app:B}

The angular distribution for a spin \decay{1/2}{1/2\; 0} baryon decay is given
by \equationname~\eqref{eq:AngDist}. The parameter $\alpha$ characterizes the parity violation
in the decay and determines the sensitivity to the initial polarization.
The effective $\alpha$ parameter for \decay{\Lc}{\Kstarzb (\Km \pip) \proton}, \decay{\Lc}{\Delta^{++} \left(\proton \pip\right) \Km} and \decay{\Lc}{\Lambda(1520)\left(\proton \Km\right) \pip}
quasi two-body decays can be calculated using the results of an 
  amplitude analysis for \decay{\Lc}{\proton\Km\pip} decays reported in Ref.~\cite{Aitala:1999uq}.
 The angular distribution for those decays is determined by the helicity amplitudes.
 A similar angular distribution to \equationname~\eqref{eq:AngDist} is obtained for the above
 quasi two-body decays when integrating over all the decay angles, except for
 the helicity angle of the baryon daughter of the \Lc.
%
%
 The computed $\alpha$ parameters are listed in \tablename~\ref{tab:alphas}.
\begin{table}[htb]
\caption{Computed $\alpha$ parameters for different quasi two-body final states in
  \decay{\Lc}{\proton\Km\pip} decays. The values for the helicity amplitudes are taken from
  Ref.~\cite{Aitala:1999uq}. Since no correlation matrix is provided in the article,
  the errors are calculated assuming no correlation among the helicity amplitude results.\label{tab:alphas}}
\centering
\begin{tabular}{lc}
\toprule
Decay & $\alpha$\\
\midrule
\decay{\Lc}{\Kstarzb (\Km \pip) \proton} & $-0.545\pm0.345$\\
\decay{\Lc}{\Delta^{++} \left(\proton \pip\right) \Km} & $-0.666\pm0.298$\\
\decay{\Lc}{\Lambda(1520)\left(\proton \Km\right) \pip} & $-0.105\pm0.604$\\
\bottomrule
\end{tabular}
\end{table}

\addcontentsline{toc}{section}{References}
\setboolean{inbibliography}{true}
\bibliographystyle{LHCb}
\bibliography{main}

\ifx\mcitethebibliography\mciteundefinedmacro
\PackageError{LHCb.bst}{mciteplus.sty has not been loaded}
{This bibstyle requires the use of the mciteplus package.}\fi
\providecommand{\href}[2]{#2}
\begin{mcitethebibliography}{10}
\mciteSetBstSublistMode{n}
\mciteSetBstMaxWidthForm{subitem}{\alph{mcitesubitemcount})}
\mciteSetBstSublistLabelBeginEnd{\mcitemaxwidthsubitemform\space}
{\relax}{\relax}

\bibitem{Purcell:1950zz}
E.~M. Purcell and N.~F. Ramsey, \ifthenelse{\boolean{articletitles}}{\emph{{On
  the Possibility of Electric Dipole Moments for Elementary Particles and
  Nuclei}}, }{}\href{http://dx.doi.org/10.1103/PhysRev.78.807}{Phys.\ Rev.\
  \textbf{78} (1950) 807}\relax
\mciteBstWouldAddEndPuncttrue
\mciteSetBstMidEndSepPunct{\mcitedefaultmidpunct}
{\mcitedefaultendpunct}{\mcitedefaultseppunct}\relax
\EndOfBibitem
\bibitem{Smith:1957ht}
J.~H. Smith, E.~M. Purcell, and N.~F. Ramsey,
  \ifthenelse{\boolean{articletitles}}{\emph{{Experimental limit to the
  electric dipole moment of the neutron}},
  }{}\href{http://dx.doi.org/10.1103/PhysRev.108.120}{Phys.\ Rev.\
  \textbf{108} (1957) 120}\relax
\mciteBstWouldAddEndPuncttrue
\mciteSetBstMidEndSepPunct{\mcitedefaultmidpunct}
{\mcitedefaultendpunct}{\mcitedefaultseppunct}\relax
\EndOfBibitem
\bibitem{Baron:2013eja}
ACME collaboration, J.~Baron {\em et~al.},
  \ifthenelse{\boolean{articletitles}}{\emph{{Order of Magnitude Smaller Limit
  on the Electric Dipole Moment of the Electron}},
  }{}\href{http://dx.doi.org/10.1126/science.1248213}{Science \textbf{343}
  (2014) 269}, \href{http://arxiv.org/abs/1310.7534}{{\normalfont\ttfamily
  arXiv:1310.7534}}\relax
\mciteBstWouldAddEndPuncttrue
\mciteSetBstMidEndSepPunct{\mcitedefaultmidpunct}
{\mcitedefaultendpunct}{\mcitedefaultseppunct}\relax
\EndOfBibitem
\bibitem{Bennett:2008dy}
Muon g-2 collaboration, G.~W. Bennett {\em et~al.},
  \ifthenelse{\boolean{articletitles}}{\emph{{An Improved Limit on the Muon
  Electric Dipole Moment}},
  }{}\href{http://dx.doi.org/10.1103/PhysRevD.80.052008}{Phys.\ Rev.\
  \textbf{D80} (2009) 052008},
  \href{http://arxiv.org/abs/0811.1207}{{\normalfont\ttfamily
  arXiv:0811.1207}}\relax
\mciteBstWouldAddEndPuncttrue
\mciteSetBstMidEndSepPunct{\mcitedefaultmidpunct}
{\mcitedefaultendpunct}{\mcitedefaultseppunct}\relax
\EndOfBibitem
\bibitem{Inami:2002ah}
Belle collaboration, K.~Inami {\em et~al.},
  \ifthenelse{\boolean{articletitles}}{\emph{{Search for the electric dipole
  moment of the tau lepton}},
  }{}\href{http://dx.doi.org/10.1016/S0370-2693(02)02984-2}{Phys.\ Lett.\
  \textbf{B551} (2003) 16},
  \href{http://arxiv.org/abs/hep-ex/0210066}{{\normalfont\ttfamily
  arXiv:hep-ex/0210066}}\relax
\mciteBstWouldAddEndPuncttrue
\mciteSetBstMidEndSepPunct{\mcitedefaultmidpunct}
{\mcitedefaultendpunct}{\mcitedefaultseppunct}\relax
\EndOfBibitem
\bibitem{Afach:2015sja}
J.~M. Pendlebury {\em et~al.},
  \ifthenelse{\boolean{articletitles}}{\emph{{Revised experimental upper limit
  on the electric dipole moment of the neutron}},
  }{}\href{http://dx.doi.org/10.1103/PhysRevD.92.092003}{Phys.\ Rev.\
  \textbf{D92} (2015), no.~9 092003},
  \href{http://arxiv.org/abs/1509.04411}{{\normalfont\ttfamily
  arXiv:1509.04411}}\relax
\mciteBstWouldAddEndPuncttrue
\mciteSetBstMidEndSepPunct{\mcitedefaultmidpunct}
{\mcitedefaultendpunct}{\mcitedefaultseppunct}\relax
\EndOfBibitem
\bibitem{Griffith:2009zz}
W.~C. Griffith {\em et~al.},
  \ifthenelse{\boolean{articletitles}}{\emph{{Improved Limit on the Permanent
  Electric Dipole Moment of Hg-199}},
  }{}\href{http://dx.doi.org/10.1103/PhysRevLett.102.101601}{Phys.\ Rev.\
  Lett.\  \textbf{102} (2009) 101601}\relax
\mciteBstWouldAddEndPuncttrue
\mciteSetBstMidEndSepPunct{\mcitedefaultmidpunct}
{\mcitedefaultendpunct}{\mcitedefaultseppunct}\relax
\EndOfBibitem
\bibitem{Dmitriev:2003sc}
V.~F. Dmitriev and R.~A. Sen'kov,
  \ifthenelse{\boolean{articletitles}}{\emph{{Schiff moment of the mercury
  nucleus and the proton dipole moment}},
  }{}\href{http://dx.doi.org/10.1103/PhysRevLett.91.212303}{Phys.\ Rev.\ Lett.\
   \textbf{91} (2003) 212303},
  \href{http://arxiv.org/abs/nucl-th/0306050}{{\normalfont\ttfamily
  arXiv:nucl-th/0306050}}\relax
\mciteBstWouldAddEndPuncttrue
\mciteSetBstMidEndSepPunct{\mcitedefaultmidpunct}
{\mcitedefaultendpunct}{\mcitedefaultseppunct}\relax
\EndOfBibitem
\bibitem{Pondrom:1981gu}
L.~Pondrom {\em et~al.}, \ifthenelse{\boolean{articletitles}}{\emph{{New Limit
  on the Electric Dipole Moment of the $\Lambda$ Hyperon}},
  }{}\href{http://dx.doi.org/10.1103/PhysRevD.23.814}{Phys.\ Rev.\
  \textbf{D23} (1981) 814}\relax
\mciteBstWouldAddEndPuncttrue
\mciteSetBstMidEndSepPunct{\mcitedefaultmidpunct}
{\mcitedefaultendpunct}{\mcitedefaultseppunct}\relax
\EndOfBibitem
\bibitem{Grange:2015fou}
Muon g-2 collaboration, J.~Grange {\em et~al.},
  \ifthenelse{\boolean{articletitles}}{\emph{{Muon (g-2) Technical Design
  Report}}, }{} tech. rep., 2015\relax
\mciteBstWouldAddEndPuncttrue
\mciteSetBstMidEndSepPunct{\mcitedefaultmidpunct}
{\mcitedefaultendpunct}{\mcitedefaultseppunct}\relax
\EndOfBibitem
\bibitem{Saito:2012zz}
J-PARC g-2/EDM collaboration, N.~Saito,
  \ifthenelse{\boolean{articletitles}}{\emph{{A novel precision measurement of
  muon g-2 and EDM at J-PARC}},
  }{}\href{http://dx.doi.org/10.1063/1.4742078}{AIP Conf.\ Proc.\
  \textbf{1467} (2012) 45}\relax
\mciteBstWouldAddEndPuncttrue
\mciteSetBstMidEndSepPunct{\mcitedefaultmidpunct}
{\mcitedefaultendpunct}{\mcitedefaultseppunct}\relax
\EndOfBibitem
\bibitem{Anastassopoulos:2015ura}
V.~Anastassopoulos {\em et~al.}, \ifthenelse{\boolean{articletitles}}{\emph{{A
  Storage Ring Experiment to Detect a Proton Electric Dipole Moment}},
  }{}\href{http://arxiv.org/abs/1502.04317}{{\normalfont\ttfamily
  arXiv:1502.04317}}, 2015\relax
\mciteBstWouldAddEndPuncttrue
\mciteSetBstMidEndSepPunct{\mcitedefaultmidpunct}
{\mcitedefaultendpunct}{\mcitedefaultseppunct}\relax
\EndOfBibitem
\bibitem{Pretz2015JEDI}
JEDI collaboration, J.~Pretz,
  \ifthenelse{\boolean{articletitles}}{\emph{Measurement of electric dipole
  moments at storage rings},
  }{}\href{http://dx.doi.org/10.1088/0031-8949/2015/T166/014035}{Physica
  Scripta \textbf{2015} (2015), no.~T166 014035}\relax
\mciteBstWouldAddEndPuncttrue
\mciteSetBstMidEndSepPunct{\mcitedefaultmidpunct}
{\mcitedefaultendpunct}{\mcitedefaultseppunct}\relax
\EndOfBibitem
\bibitem{Khriplovich:1998zq}
I.~B. Khriplovich, \ifthenelse{\boolean{articletitles}}{\emph{{Feasibility of
  search for nuclear electric dipole moments at ion storage rings}},
  }{}\href{http://dx.doi.org/10.1016/S0370-2693(98)01353-7}{Phys.\ Lett.\
  \textbf{B444} (1998) 98},
  \href{http://arxiv.org/abs/hep-ph/9809336}{{\normalfont\ttfamily
  arXiv:hep-ph/9809336}}\relax
\mciteBstWouldAddEndPuncttrue
\mciteSetBstMidEndSepPunct{\mcitedefaultmidpunct}
{\mcitedefaultendpunct}{\mcitedefaultseppunct}\relax
\EndOfBibitem
\bibitem{Engel:2013lsa}
J.~Engel, M.~J. Ramsey-Musolf, and U.~van Kolck,
  \ifthenelse{\boolean{articletitles}}{\emph{{Electric Dipole Moments of
  Nucleons, Nuclei, and Atoms: The Standard Model and Beyond}},
  }{}\href{http://dx.doi.org/10.1016/j.ppnp.2013.03.003}{Prog.\ Part.\ Nucl.\
  Phys.\  \textbf{71} (2013) 21},
  \href{http://arxiv.org/abs/1303.2371}{{\normalfont\ttfamily
  arXiv:1303.2371}}\relax
\mciteBstWouldAddEndPuncttrue
\mciteSetBstMidEndSepPunct{\mcitedefaultmidpunct}
{\mcitedefaultendpunct}{\mcitedefaultseppunct}\relax
\EndOfBibitem
\bibitem{Fukuyama:2012np}
T.~Fukuyama, \ifthenelse{\boolean{articletitles}}{\emph{{Searching for New
  Physics beyond the Standard Model in Electric Dipole Moment}},
  }{}\href{http://dx.doi.org/10.1142/S0217751X12300153}{Int.\ J.\ Mod.\ Phys.\
  \textbf{A27} (2012) 1230015},
  \href{http://arxiv.org/abs/1201.4252}{{\normalfont\ttfamily
  arXiv:1201.4252}}\relax
\mciteBstWouldAddEndPuncttrue
\mciteSetBstMidEndSepPunct{\mcitedefaultmidpunct}
{\mcitedefaultendpunct}{\mcitedefaultseppunct}\relax
\EndOfBibitem
\bibitem{Jungmann:2013sga}
K.~Jungmann, \ifthenelse{\boolean{articletitles}}{\emph{{Searching for electric
  dipole moments}}, }{}\href{http://dx.doi.org/10.1002/andp.201300071}{Annalen
  Phys.\  \textbf{525} (2013), no.~8-9 550}\relax
\mciteBstWouldAddEndPuncttrue
\mciteSetBstMidEndSepPunct{\mcitedefaultmidpunct}
{\mcitedefaultendpunct}{\mcitedefaultseppunct}\relax
\EndOfBibitem
\bibitem{Pospelov:2005pr}
M.~Pospelov and A.~Ritz, \ifthenelse{\boolean{articletitles}}{\emph{{Electric
  dipole moments as probes of new physics}},
  }{}\href{http://dx.doi.org/10.1016/j.aop.2005.04.002}{Annals Phys.\
  \textbf{318} (2005) 119},
  \href{http://arxiv.org/abs/hep-ph/0504231}{{\normalfont\ttfamily
  arXiv:hep-ph/0504231}}\relax
\mciteBstWouldAddEndPuncttrue
\mciteSetBstMidEndSepPunct{\mcitedefaultmidpunct}
{\mcitedefaultendpunct}{\mcitedefaultseppunct}\relax
\EndOfBibitem
\bibitem{Semertzidis:2011zz}
Y.~K. Semertzidis, \ifthenelse{\boolean{articletitles}}{\emph{{Review of EDM
  experiments}}, }{}\href{http://dx.doi.org/10.1088/1742-6596/335/1/012012}{J.\
  Phys.\ Conf.\ Ser.\  \textbf{335} (2011) 335:012012}\relax
\mciteBstWouldAddEndPuncttrue
\mciteSetBstMidEndSepPunct{\mcitedefaultmidpunct}
{\mcitedefaultendpunct}{\mcitedefaultseppunct}\relax
\EndOfBibitem
\bibitem{Semertzidis:2016wtd}
Y.~K. Semertzidis, \ifthenelse{\boolean{articletitles}}{\emph{{Storage Ring EDM
  Experiments}}, }{}\href{http://dx.doi.org/10.1051/epjconf/201611801032}{EPJ
  Web Conf.\  \textbf{118} (2016) 01032}\relax
\mciteBstWouldAddEndPuncttrue
\mciteSetBstMidEndSepPunct{\mcitedefaultmidpunct}
{\mcitedefaultendpunct}{\mcitedefaultseppunct}\relax
\EndOfBibitem
\bibitem{Onderwater2011}
C.~J.~G. Onderwater, \ifthenelse{\boolean{articletitles}}{\emph{{Search for
  EDMs using storage rings}},
  }{}\href{http://dx.doi.org/10.1088/1742-6596/295/1/012008}{J.\ Phys.\ Conf.\
  Ser \textbf{295} (2011), no.~1 012008}\relax
\mciteBstWouldAddEndPuncttrue
\mciteSetBstMidEndSepPunct{\mcitedefaultmidpunct}
{\mcitedefaultendpunct}{\mcitedefaultseppunct}\relax
\EndOfBibitem
\bibitem{Peccei:1977hh}
R.~D. Peccei and H.~R. Quinn, \ifthenelse{\boolean{articletitles}}{\emph{{\CP
  Conservation in the Presence of Instantons}},
  }{}\href{http://dx.doi.org/10.1103/PhysRevLett.38.1440}{Phys.\ Rev.\ Lett.\
  \textbf{38} (1977) 1440}\relax
\mciteBstWouldAddEndPuncttrue
\mciteSetBstMidEndSepPunct{\mcitedefaultmidpunct}
{\mcitedefaultendpunct}{\mcitedefaultseppunct}\relax
\EndOfBibitem
\bibitem{Weinberg:1977ma}
S.~Weinberg, \ifthenelse{\boolean{articletitles}}{\emph{{A New Light Boson?}},
  }{}\href{http://dx.doi.org/10.1103/PhysRevLett.40.223}{Phys.\ Rev.\ Lett.\
  \textbf{40} (1978) 223}\relax
\mciteBstWouldAddEndPuncttrue
\mciteSetBstMidEndSepPunct{\mcitedefaultmidpunct}
{\mcitedefaultendpunct}{\mcitedefaultseppunct}\relax
\EndOfBibitem
\bibitem{Wilczek:1977pj}
F.~Wilczek, \ifthenelse{\boolean{articletitles}}{\emph{{Problem of Strong P and
  T Invariance in the Presence of Instantons}},
  }{}\href{http://dx.doi.org/10.1103/PhysRevLett.40.279}{Phys.\ Rev.\ Lett.\
  \textbf{40} (1978) 279}\relax
\mciteBstWouldAddEndPuncttrue
\mciteSetBstMidEndSepPunct{\mcitedefaultmidpunct}
{\mcitedefaultendpunct}{\mcitedefaultseppunct}\relax
\EndOfBibitem
\bibitem{Guo:2012vf}
F.-K. Guo and U.-G. Meissner,
  \ifthenelse{\boolean{articletitles}}{\emph{{Baryon electric dipole moments
  from strong \CP violation}},
  }{}\href{http://dx.doi.org/10.1007/JHEP12(2012)097}{JHEP \textbf{12} (2012)
  097}, \href{http://arxiv.org/abs/1210.5887}{{\normalfont\ttfamily
  arXiv:1210.5887}}\relax
\mciteBstWouldAddEndPuncttrue
\mciteSetBstMidEndSepPunct{\mcitedefaultmidpunct}
{\mcitedefaultendpunct}{\mcitedefaultseppunct}\relax
\EndOfBibitem
\bibitem{Atwood:1992fb}
D.~Atwood and A.~Soni, \ifthenelse{\boolean{articletitles}}{\emph{{Chiral
  perturbation theory constraint on the electric dipole moment of the $\Lambda$
  hyperon}}, }{}\href{http://dx.doi.org/10.1016/0370-2693(92)91048-E}{Phys.\
  Lett.\  \textbf{B291} (1992) 293}\relax
\mciteBstWouldAddEndPuncttrue
\mciteSetBstMidEndSepPunct{\mcitedefaultmidpunct}
{\mcitedefaultendpunct}{\mcitedefaultseppunct}\relax
\EndOfBibitem
\bibitem{Pich:1991fq}
A.~Pich and E.~de~Rafael, \ifthenelse{\boolean{articletitles}}{\emph{{Strong CP
  violation in an effective chiral Lagrangian approach}},
  }{}\href{http://dx.doi.org/10.1016/0550-3213(91)90019-T}{Nucl.\ Phys.\
  \textbf{B367} (1991) 313}\relax
\mciteBstWouldAddEndPuncttrue
\mciteSetBstMidEndSepPunct{\mcitedefaultmidpunct}
{\mcitedefaultendpunct}{\mcitedefaultseppunct}\relax
\EndOfBibitem
\bibitem{Borasoy:2000pq}
B.~Borasoy, \ifthenelse{\boolean{articletitles}}{\emph{{The electric dipole
  moment of the neutron in chiral perturbation theory}},
  }{}\href{http://dx.doi.org/10.1103/PhysRevD.61.114017}{Phys.\ Rev.\
  \textbf{D61} (2000) 114017},
  \href{http://arxiv.org/abs/hep-ph/0004011}{{\normalfont\ttfamily
  arXiv:hep-ph/0004011}}\relax
\mciteBstWouldAddEndPuncttrue
\mciteSetBstMidEndSepPunct{\mcitedefaultmidpunct}
{\mcitedefaultendpunct}{\mcitedefaultseppunct}\relax
\EndOfBibitem
\bibitem{Sala:2013osa}
F.~Sala, \ifthenelse{\boolean{articletitles}}{\emph{{A bound on the charm
  chromo-EDM and its implications}},
  }{}\href{http://dx.doi.org/10.1007/JHEP03(2014)061}{JHEP \textbf{03} (2014)
  061}, \href{http://arxiv.org/abs/1312.2589}{{\normalfont\ttfamily
  arXiv:1312.2589}}\relax
\mciteBstWouldAddEndPuncttrue
\mciteSetBstMidEndSepPunct{\mcitedefaultmidpunct}
{\mcitedefaultendpunct}{\mcitedefaultseppunct}\relax
\EndOfBibitem
\bibitem{DiSciacca:2013hya}
ATRAP collaboration, J.~DiSciacca {\em et~al.},
  \ifthenelse{\boolean{articletitles}}{\emph{{One-Particle Measurement of the
  Antiproton Magnetic Moment}},
  }{}\href{http://dx.doi.org/10.1103/PhysRevLett.110.130801}{Phys.\ Rev.\
  Lett.\  \textbf{110} (2013), no.~13 130801},
  \href{http://arxiv.org/abs/1301.6310}{{\normalfont\ttfamily
  arXiv:1301.6310}}\relax
\mciteBstWouldAddEndPuncttrue
\mciteSetBstMidEndSepPunct{\mcitedefaultmidpunct}
{\mcitedefaultendpunct}{\mcitedefaultseppunct}\relax
\EndOfBibitem
\bibitem{VanDyck:1987ay}
R.~S. Van~Dyck, P.~B. Schwinberg, and H.~G. Dehmelt,
  \ifthenelse{\boolean{articletitles}}{\emph{{New high-precision comparison of
  electron and positron \textit{g} factors}},
  }{}\href{http://dx.doi.org/10.1103/PhysRevLett.59.26}{Phys.\ Rev.\ Lett.\
  \textbf{59} (1987) 26}\relax
\mciteBstWouldAddEndPuncttrue
\mciteSetBstMidEndSepPunct{\mcitedefaultmidpunct}
{\mcitedefaultendpunct}{\mcitedefaultseppunct}\relax
\EndOfBibitem
\bibitem{Bennett:2004pv}
Muon g-2 collaboration, G.~W. Bennett {\em et~al.},
  \ifthenelse{\boolean{articletitles}}{\emph{{Measurement of the negative muon
  anomalous magnetic moment to 0.7 ppm}},
  }{}\href{http://dx.doi.org/10.1103/PhysRevLett.92.161802}{Phys.\ Rev.\ Lett.\
   \textbf{92} (2004) 161802},
  \href{http://arxiv.org/abs/hep-ex/0401008}{{\normalfont\ttfamily
  arXiv:hep-ex/0401008}}\relax
\mciteBstWouldAddEndPuncttrue
\mciteSetBstMidEndSepPunct{\mcitedefaultmidpunct}
{\mcitedefaultendpunct}{\mcitedefaultseppunct}\relax
\EndOfBibitem
\bibitem{Ulmer:2013rra}
S.~Ulmer {\em et~al.}, \ifthenelse{\boolean{articletitles}}{\emph{{Technical
  Design Report BASE}}, }{} Tech. Rep. CERN-SPSC-2013-002. SPSC-TDR-002,
  2013\relax
\mciteBstWouldAddEndPuncttrue
\mciteSetBstMidEndSepPunct{\mcitedefaultmidpunct}
{\mcitedefaultendpunct}{\mcitedefaultseppunct}\relax
\EndOfBibitem
\bibitem{Baryshevsky:2016cul}
V.~G. Baryshevsky, \ifthenelse{\boolean{articletitles}}{\emph{{The possibility
  to measure the magnetic moments of short-lived particles (charm and beauty
  baryons) at LHC and FCC energies using the phenomenon of spin rotation in
  crystals}}, }{}\href{http://dx.doi.org/10.1016/j.physletb.2016.04.025}{Phys.\
  Lett.\  \textbf{B757} (2016) 426}\relax
\mciteBstWouldAddEndPuncttrue
\mciteSetBstMidEndSepPunct{\mcitedefaultmidpunct}
{\mcitedefaultendpunct}{\mcitedefaultseppunct}\relax
\EndOfBibitem
\bibitem{Burmistrov:2194564}
UA9 collaboration, L.~Burmistrov {\em et~al.},
  \ifthenelse{\boolean{articletitles}}{\emph{{Measurement of Short Living
  Baryon Magnetic Moment using Bent Crystals at SPS and LHC}}, }{} Tech. Rep.
  CERN-SPSC-2016-030. SPSC-EOI-012, CERN, Geneva, Jun, 2016\relax
\mciteBstWouldAddEndPuncttrue
\mciteSetBstMidEndSepPunct{\mcitedefaultmidpunct}
{\mcitedefaultendpunct}{\mcitedefaultseppunct}\relax
\EndOfBibitem
\bibitem{Heller:1978ty}
K.~J. Heller {\em et~al.},
  \ifthenelse{\boolean{articletitles}}{\emph{{Polarization of
  $\ensuremath{\Lambda}'\mathrm{s}$ and
  $\overline{\mathrm{\ensuremath{\Lambda}}}'\mathrm{s}$ Produced by 400-GeV
  Protons}}, }{}\href{http://dx.doi.org/10.1103/PhysRevLett.41.607}{Phys.\
  Rev.\ Lett.\  \textbf{41} (1978) 607}, [Erratum: Phys. Rev.
  Lett.45,1043(1980)]\relax
\mciteBstWouldAddEndPuncttrue
\mciteSetBstMidEndSepPunct{\mcitedefaultmidpunct}
{\mcitedefaultendpunct}{\mcitedefaultseppunct}\relax
\EndOfBibitem
\bibitem{Link:2005ft}
FOCUS collaboration, J.~M. Link {\em et~al.},
  \ifthenelse{\boolean{articletitles}}{\emph{{Study of the decay asymmetry
  parameter and CP violation parameter in the $\Lambda_c^+ \to \Lambda \pi^+$
  decay}}, }{}\href{http://dx.doi.org/10.1016/j.physletb.2006.01.017}{Phys.\
  Lett.\  \textbf{B634} (2006) 165},
  \href{http://arxiv.org/abs/hep-ex/0509042}{{\normalfont\ttfamily
  arXiv:hep-ex/0509042}}\relax
\mciteBstWouldAddEndPuncttrue
\mciteSetBstMidEndSepPunct{\mcitedefaultmidpunct}
{\mcitedefaultendpunct}{\mcitedefaultseppunct}\relax
\EndOfBibitem
\bibitem{Olive:2016xmw}
C.~Patrignani, \ifthenelse{\boolean{articletitles}}{\emph{{Review of Particle
  Physics}}, }{}\href{http://dx.doi.org/10.1088/1674-1137/40/10/100001}{Chin.\
  Phys.\  \textbf{C40} (2016), no.~10 100001}\relax
\mciteBstWouldAddEndPuncttrue
\mciteSetBstMidEndSepPunct{\mcitedefaultmidpunct}
{\mcitedefaultendpunct}{\mcitedefaultseppunct}\relax
\EndOfBibitem
\bibitem{Aaij:2013oxa}
LHCb collaboration, R.~Aaij {\em et~al.},
  \ifthenelse{\boolean{articletitles}}{\emph{{Measurements of the $\Lambda_b^0
  \to J/\psi \Lambda$ decay amplitudes and the $\Lambda_b^0$ polarisation in
  $pp$ collisions at $\sqrt{s} = 7$ TeV}},
  }{}\href{http://dx.doi.org/10.1016/j.physletb.2013.05.041}{Phys.\ Lett.\
  \textbf{B724} (2013) 27},
  \href{http://arxiv.org/abs/1302.5578}{{\normalfont\ttfamily
  arXiv:1302.5578}}\relax
\mciteBstWouldAddEndPuncttrue
\mciteSetBstMidEndSepPunct{\mcitedefaultmidpunct}
{\mcitedefaultendpunct}{\mcitedefaultseppunct}\relax
\EndOfBibitem
\bibitem{Aad:2014iba}
ATLAS collaboration, G.~Aad {\em et~al.},
  \ifthenelse{\boolean{articletitles}}{\emph{{Measurement of the
  parity-violating asymmetry parameter $\alpha_b$ and the helicity amplitudes
  for the decay $\Lambda_b^0\to J/\psi \Lambda^0$ with the ATLAS detector}},
  }{}\href{http://dx.doi.org/10.1103/PhysRevD.89.092009}{Phys.\ Rev.\
  \textbf{D89} (2014), no.~9 092009},
  \href{http://arxiv.org/abs/1404.1071}{{\normalfont\ttfamily
  arXiv:1404.1071}}\relax
\mciteBstWouldAddEndPuncttrue
\mciteSetBstMidEndSepPunct{\mcitedefaultmidpunct}
{\mcitedefaultendpunct}{\mcitedefaultseppunct}\relax
\EndOfBibitem
\bibitem{Lee:1957qs}
T.~D. Lee and C.-N. Yang, \ifthenelse{\boolean{articletitles}}{\emph{{General
  Partial Wave Analysis of the Decay of a Hyperon of Spin 1/2}},
  }{}\href{http://dx.doi.org/10.1103/PhysRev.108.1645}{Phys.\ Rev.\
  \textbf{108} (1957) 1645}\relax
\mciteBstWouldAddEndPuncttrue
\mciteSetBstMidEndSepPunct{\mcitedefaultmidpunct}
{\mcitedefaultendpunct}{\mcitedefaultseppunct}\relax
\EndOfBibitem
\bibitem{Richman1984}
J.~D. Richman, \ifthenelse{\boolean{articletitles}}{\emph{{An experimenter's
  guide to the helicity formalism}}, }{} Tech. Rep. CALT-68-1148, Calif. Inst.
  Technol., Pasadena, CA, Jun, 1984\relax
\mciteBstWouldAddEndPuncttrue
\mciteSetBstMidEndSepPunct{\mcitedefaultmidpunct}
{\mcitedefaultendpunct}{\mcitedefaultseppunct}\relax
\EndOfBibitem
\bibitem{Thomas:1926dy}
L.~H. Thomas, \ifthenelse{\boolean{articletitles}}{\emph{{The motion of a
  spinning electron}}, }{}\href{http://dx.doi.org/10.1038/117514a0}{Nature
  \textbf{117} (1926) 514}\relax
\mciteBstWouldAddEndPuncttrue
\mciteSetBstMidEndSepPunct{\mcitedefaultmidpunct}
{\mcitedefaultendpunct}{\mcitedefaultseppunct}\relax
\EndOfBibitem
\bibitem{Thomas:1927yu}
L.~H. Thomas, \ifthenelse{\boolean{articletitles}}{\emph{{The kinematics of an
  electron with an axis}}, }{}Phil.\ Mag.\  \textbf{3} (1927) 1\relax
\mciteBstWouldAddEndPuncttrue
\mciteSetBstMidEndSepPunct{\mcitedefaultmidpunct}
{\mcitedefaultendpunct}{\mcitedefaultseppunct}\relax
\EndOfBibitem
\bibitem{Bargmann:1959gz}
V.~Bargmann, L.~Michel, and V.~L. Telegdi,
  \ifthenelse{\boolean{articletitles}}{\emph{Precession of the polarization of
  particles moving in a homogeneous electromagnetic field},
  }{}\href{http://dx.doi.org/10.1103/PhysRevLett.2.435}{Phys.\ Rev.\ Lett.\
  \textbf{2} (1959) 435}\relax
\mciteBstWouldAddEndPuncttrue
\mciteSetBstMidEndSepPunct{\mcitedefaultmidpunct}
{\mcitedefaultendpunct}{\mcitedefaultseppunct}\relax
\EndOfBibitem
\bibitem{LHCb-DP-2014-002}
LHCb collaboration, R.~Aaij {\em et~al.},
  \ifthenelse{\boolean{articletitles}}{\emph{{LHCb detector performance}},
  }{}\href{http://dx.doi.org/10.1142/S0217751X15300227}{Int.\ J.\ Mod.\ Phys.\
  \textbf{A30} (2015) 1530022},
  \href{http://arxiv.org/abs/1412.6352}{{\normalfont\ttfamily
  arXiv:1412.6352}}\relax
\mciteBstWouldAddEndPuncttrue
\mciteSetBstMidEndSepPunct{\mcitedefaultmidpunct}
{\mcitedefaultendpunct}{\mcitedefaultseppunct}\relax
\EndOfBibitem
\bibitem{Schachinger:1978qs}
L.~Schachinger {\em et~al.}, \ifthenelse{\boolean{articletitles}}{\emph{{A
  Precise Measurement of the $\Lambda^0$ Magnetic Moment}},
  }{}\href{http://dx.doi.org/10.1103/PhysRevLett.41.1348}{Phys.\ Rev.\ Lett.\
  \textbf{41} (1978) 1348}\relax
\mciteBstWouldAddEndPuncttrue
\mciteSetBstMidEndSepPunct{\mcitedefaultmidpunct}
{\mcitedefaultendpunct}{\mcitedefaultseppunct}\relax
\EndOfBibitem
\bibitem{Chen:1992wx}
E761 collaboration, D.~Chen {\em et~al.},
  \ifthenelse{\boolean{articletitles}}{\emph{{First observation of magnetic
  moment precession of channeled particles in bent crystals}},
  }{}\href{http://dx.doi.org/10.1103/PhysRevLett.69.3286}{Phys.\ Rev.\ Lett.\
  \textbf{69} (1992) 3286}\relax
\mciteBstWouldAddEndPuncttrue
\mciteSetBstMidEndSepPunct{\mcitedefaultmidpunct}
{\mcitedefaultendpunct}{\mcitedefaultseppunct}\relax
\EndOfBibitem
\bibitem{Baublis:1994ku}
V.~V. Baublis {\em et~al.},
  \ifthenelse{\boolean{articletitles}}{\emph{{Measuring the magnetic moments of
  short-lived particles using channeling in bent crystals}},
  }{}\href{http://dx.doi.org/10.1016/0168-583X(94)95524-7}{Nucl.\ Instrum.\
  Meth.\  \textbf{B90} (1994) 112}\relax
\mciteBstWouldAddEndPuncttrue
\mciteSetBstMidEndSepPunct{\mcitedefaultmidpunct}
{\mcitedefaultendpunct}{\mcitedefaultseppunct}\relax
\EndOfBibitem
\bibitem{Samsonov:1996ah}
V.~M. Samsonov, \ifthenelse{\boolean{articletitles}}{\emph{{On the possibility
  of measuring charm baryon magnetic moments with channeling}},
  }{}\href{http://dx.doi.org/10.1016/0168-583X(96)00348-5}{Nucl.\ Instrum.\
  Meth.\  \textbf{B119} (1996) 271}\relax
\mciteBstWouldAddEndPuncttrue
\mciteSetBstMidEndSepPunct{\mcitedefaultmidpunct}
{\mcitedefaultendpunct}{\mcitedefaultseppunct}\relax
\EndOfBibitem
\bibitem{Jacob:1959at}
M.~Jacob and G.~C. Wick, \ifthenelse{\boolean{articletitles}}{\emph{{On the
  general theory of collisions for particles with spin}},
  }{}\href{http://dx.doi.org/10.1016/0003-4916(59)90051-X}{Annals Phys.\
  \textbf{7} (1959) 404}\relax
\mciteBstWouldAddEndPuncttrue
\mciteSetBstMidEndSepPunct{\mcitedefaultmidpunct}
{\mcitedefaultendpunct}{\mcitedefaultseppunct}\relax
\EndOfBibitem
\bibitem{Baryshevsky:2015zba}
V.~G. Baryshevsky, \ifthenelse{\boolean{articletitles}}{\emph{{Spin rotation
  and depolarization of high-energy particles in crystals at Hadron Collider
  (LHC) and Future Circular Collider (FCC) energies and the possibility to
  measure the anomalous magnetic moments of short-lived particles}},
  }{}\href{http://arxiv.org/abs/1504.06702}{{\normalfont\ttfamily
  arXiv:1504.06702}}, 2015\relax
\mciteBstWouldAddEndPuncttrue
\mciteSetBstMidEndSepPunct{\mcitedefaultmidpunct}
{\mcitedefaultendpunct}{\mcitedefaultseppunct}\relax
\EndOfBibitem
\bibitem{Kim:1982ry}
I.~J. Kim, \ifthenelse{\boolean{articletitles}}{\emph{{Magnetic moment
  measurement of baryons with heavy flavored quarks by planar channeling
  through bent crystal}},
  }{}\href{http://dx.doi.org/10.1016/0550-3213(83)90363-2}{Nucl.\ Phys.\
  \textbf{B229} (1983) 251}\relax
\mciteBstWouldAddEndPuncttrue
\mciteSetBstMidEndSepPunct{\mcitedefaultmidpunct}
{\mcitedefaultendpunct}{\mcitedefaultseppunct}\relax
\EndOfBibitem
\bibitem{Lyuboshits:1979qw}
V.~L. Lyuboshits, \ifthenelse{\boolean{articletitles}}{\emph{{The Spin Rotation
  at Deflection of Relativistic Charged Particle in Electric Field}}, }{}Sov.\
  J.\ Nucl.\ Phys.\  \textbf{31} (1980) 509\relax
\mciteBstWouldAddEndPuncttrue
\mciteSetBstMidEndSepPunct{\mcitedefaultmidpunct}
{\mcitedefaultendpunct}{\mcitedefaultseppunct}\relax
\EndOfBibitem
\bibitem{Aitala:1999uq}
E791 collaboration, E.~M. Aitala {\em et~al.},
  \ifthenelse{\boolean{articletitles}}{\emph{{Multidimensional resonance
  analysis of $\Lc \to p K^- \pi^+$}},
  }{}\href{http://dx.doi.org/10.1016/S0370-2693(99)01397-0}{Phys.\ Lett.\
  \textbf{B471} (2000) 449},
  \href{http://arxiv.org/abs/hep-ex/9912003}{{\normalfont\ttfamily
  arXiv:hep-ex/9912003}}\relax
\mciteBstWouldAddEndPuncttrue
\mciteSetBstMidEndSepPunct{\mcitedefaultmidpunct}
{\mcitedefaultendpunct}{\mcitedefaultseppunct}\relax
\EndOfBibitem
\bibitem{Szwed:1981rr}
J.~Szwed, \ifthenelse{\boolean{articletitles}}{\emph{{Hyperon Polarization at
  High-Energies}},
  }{}\href{http://dx.doi.org/10.1016/0370-2693(81)90788-7}{Phys.\ Lett.\
  \textbf{B105} (1981) 403}\relax
\mciteBstWouldAddEndPuncttrue
\mciteSetBstMidEndSepPunct{\mcitedefaultmidpunct}
{\mcitedefaultendpunct}{\mcitedefaultseppunct}\relax
\EndOfBibitem
\bibitem{Jezabek:1992ke}
M.~Jezabek, K.~Rybicki, and R.~Rylko,
  \ifthenelse{\boolean{articletitles}}{\emph{{Experimental study of spin
  effects in hadroproduction and decay of $\Lambda_c^+$}},
  }{}\href{http://dx.doi.org/10.1016/0370-2693(92)90177-6}{Phys.\ Lett.\
  \textbf{B286} (1992) 175}\relax
\mciteBstWouldAddEndPuncttrue
\mciteSetBstMidEndSepPunct{\mcitedefaultmidpunct}
{\mcitedefaultendpunct}{\mcitedefaultseppunct}\relax
\EndOfBibitem
\bibitem{Aaij:2015bpa}
LHCb collaboration, R.~Aaij {\em et~al.},
  \ifthenelse{\boolean{articletitles}}{\emph{{Measurements of prompt charm
  production cross-sections in $pp$ collisions at $ \sqrt{s}=13 $ TeV}},
  }{}\href{http://dx.doi.org/10.1007/JHEP03(2016)159}{JHEP \textbf{03} (2016)
  159}, \href{http://arxiv.org/abs/1510.01707}{{\normalfont\ttfamily
  arXiv:1510.01707}}, [Erratum: JHEP09,013(2016)]\relax
\mciteBstWouldAddEndPuncttrue
\mciteSetBstMidEndSepPunct{\mcitedefaultmidpunct}
{\mcitedefaultendpunct}{\mcitedefaultseppunct}\relax
\EndOfBibitem
\bibitem{FONLLWEB}
M.~Cacciari, \ifthenelse{\boolean{articletitles}}{\emph{{FONLL Heavy Quark
  Production}}, }{} {\url{http://www.lpthe.jussieu.fr/
  ~cacciari/fonll/fonllform.html}}.
\newblock {Accessed: 17.05.2016}\relax
\mciteBstWouldAddEndPuncttrue
\mciteSetBstMidEndSepPunct{\mcitedefaultmidpunct}
{\mcitedefaultendpunct}{\mcitedefaultseppunct}\relax
\EndOfBibitem
\bibitem{Aaij:2010gn}
LHCb collaboration, R.~Aaij {\em et~al.},
  \ifthenelse{\boolean{articletitles}}{\emph{{Measurement of $\sigma(pp \to b
  \bar{b} X)$ at $\sqrt{s}=7~\rm{TeV}$ in the forward region}},
  }{}\href{http://dx.doi.org/10.1016/j.physletb.2010.10.010}{Phys.\ Lett.\
  \textbf{B694} (2010) 209},
  \href{http://arxiv.org/abs/1009.2731}{{\normalfont\ttfamily
  arXiv:1009.2731}}\relax
\mciteBstWouldAddEndPuncttrue
\mciteSetBstMidEndSepPunct{\mcitedefaultmidpunct}
{\mcitedefaultendpunct}{\mcitedefaultseppunct}\relax
\EndOfBibitem
\bibitem{Aaij:2015rla}
LHCb collaboration, R.~Aaij {\em et~al.},
  \ifthenelse{\boolean{articletitles}}{\emph{{Measurement of forward $J/\psi$
  production cross-sections in $pp$ collisions at $\sqrt{s}=13$ TeV}},
  }{}\href{http://dx.doi.org/10.1007/JHEP10(2015)172}{JHEP \textbf{10} (2015)
  172}, \href{http://arxiv.org/abs/1509.00771}{{\normalfont\ttfamily
  arXiv:1509.00771}}\relax
\mciteBstWouldAddEndPuncttrue
\mciteSetBstMidEndSepPunct{\mcitedefaultmidpunct}
{\mcitedefaultendpunct}{\mcitedefaultseppunct}\relax
\EndOfBibitem
\bibitem{Lisovyi:2015uqa}
M.~Lisovyi, A.~Verbytskyi, and O.~Zenaiev,
  \ifthenelse{\boolean{articletitles}}{\emph{{Combined analysis of charm-quark
  fragmentation-fraction measurements}},
  }{}\href{http://dx.doi.org/10.1140/epjc/s10052-016-4246-y}{Eur.\ Phys.\ J.\
  \textbf{C76} (2016), no.~7 397},
  \href{http://arxiv.org/abs/1509.01061}{{\normalfont\ttfamily
  arXiv:1509.01061}}\relax
\mciteBstWouldAddEndPuncttrue
\mciteSetBstMidEndSepPunct{\mcitedefaultmidpunct}
{\mcitedefaultendpunct}{\mcitedefaultseppunct}\relax
\EndOfBibitem
\bibitem{Gladilin:2014tba}
L.~Gladilin, \ifthenelse{\boolean{articletitles}}{\emph{{Fragmentation
  fractions of $c$ and $b$ quarks into charmed hadrons at LEP}},
  }{}\href{http://dx.doi.org/10.1140/epjc/s10052-014-3250-3}{Eur.\ Phys.\ J.\
  \textbf{C75} (2015), no.~1 19},
  \href{http://arxiv.org/abs/1404.3888}{{\normalfont\ttfamily
  arXiv:1404.3888}}\relax
\mciteBstWouldAddEndPuncttrue
\mciteSetBstMidEndSepPunct{\mcitedefaultmidpunct}
{\mcitedefaultendpunct}{\mcitedefaultseppunct}\relax
\EndOfBibitem
\bibitem{Amhis:2014hma}
Heavy Flavor Averaging Group (HFAG) collaboration, Y.~Amhis {\em et~al.},
  \ifthenelse{\boolean{articletitles}}{\emph{{Averages of $b$-hadron,
  $c$-hadron, and $\tau$-lepton properties as of summer 2014}},
  }{}\href{http://arxiv.org/abs/1412.7515}{{\normalfont\ttfamily
  arXiv:1412.7515}}\relax
\mciteBstWouldAddEndPuncttrue
\mciteSetBstMidEndSepPunct{\mcitedefaultmidpunct}
{\mcitedefaultendpunct}{\mcitedefaultseppunct}\relax
\EndOfBibitem
\bibitem{Galanti:2015pqa}
M.~Galanti {\em et~al.}, \ifthenelse{\boolean{articletitles}}{\emph{{Heavy
  baryons as polarimeters at colliders}},
  }{}\href{http://dx.doi.org/10.1007/JHEP11(2015)067}{JHEP \textbf{11} (2015)
  067}, \href{http://arxiv.org/abs/1505.02771}{{\normalfont\ttfamily
  arXiv:1505.02771}}\relax
\mciteBstWouldAddEndPuncttrue
\mciteSetBstMidEndSepPunct{\mcitedefaultmidpunct}
{\mcitedefaultendpunct}{\mcitedefaultseppunct}\relax
\EndOfBibitem
\bibitem{Sjostrand:2006za}
T.~Sjostrand, S.~Mrenna, and P.~Z. Skands,
  \ifthenelse{\boolean{articletitles}}{\emph{{PYTHIA 6.4 Physics and Manual}},
  }{}\href{http://dx.doi.org/10.1088/1126-6708/2006/05/026}{JHEP \textbf{05}
  (2006) 026}, \href{http://arxiv.org/abs/hep-ph/0603175}{{\normalfont\ttfamily
  arXiv:hep-ph/0603175}}\relax
\mciteBstWouldAddEndPuncttrue
\mciteSetBstMidEndSepPunct{\mcitedefaultmidpunct}
{\mcitedefaultendpunct}{\mcitedefaultseppunct}\relax
\EndOfBibitem
\bibitem{Lange:2001uf}
D.~J. Lange, \ifthenelse{\boolean{articletitles}}{\emph{{The EvtGen particle
  decay simulation package}},
  }{}\href{http://dx.doi.org/10.1016/S0168-9002(01)00089-4}{Nucl.\ Instrum.\
  Meth.\  \textbf{A462} (2001) 152}\relax
\mciteBstWouldAddEndPuncttrue
\mciteSetBstMidEndSepPunct{\mcitedefaultmidpunct}
{\mcitedefaultendpunct}{\mcitedefaultseppunct}\relax
\EndOfBibitem
\bibitem{Hicheur:2007jfk}
A.~Hicheur and G.~Conti,
  \ifthenelse{\boolean{articletitles}}{\emph{{Parameterization of the LHCb
  magnetic field map}}, }{} in {\em {Proceedings, 2007 IEEE Nuclear Science
  Symposium and Medical Imaging Conference (NSS/MIC 2007): Honolulu, Hawaii,
  October 28-November 3, 2007}}, pp.~2439--2443, 2007.
\newblock
  doi:~\href{http://dx.doi.org/10.1109/NSSMIC.2007.4436650}{10.1109/NSSMIC.200%
7.4436650}\relax
\mciteBstWouldAddEndPuncttrue
\mciteSetBstMidEndSepPunct{\mcitedefaultmidpunct}
{\mcitedefaultendpunct}{\mcitedefaultseppunct}\relax
\EndOfBibitem
\bibitem{LHCb-TDR-016}
LHCb collaboration, \ifthenelse{\boolean{articletitles}}{\emph{{LHCb Trigger
  and Online Technical Design Report}}, }{}
  \href{http://cdsweb.cern.ch/search?p=CERN-LHCC-2014-016&f=reportnumber&actio%
n_search=Search&c=LHCb+Reports} {CERN-LHCC-2014-016}.
\newblock {LHCb-TDR-016}\relax
\mciteBstWouldAddEndPuncttrue
\mciteSetBstMidEndSepPunct{\mcitedefaultmidpunct}
{\mcitedefaultendpunct}{\mcitedefaultseppunct}\relax
\EndOfBibitem
\bibitem{LHCb-TDR-015}
LHCb collaboration, \ifthenelse{\boolean{articletitles}}{\emph{{LHCb Tracker
  Upgrade Technical Design Report}}, }{}
  \href{http://cdsweb.cern.ch/search?p=CERN-LHCC-2014-001&f=reportnumber&actio%
n_search=Search&c=LHCb+Reports} {CERN-LHCC-2014-001}.
\newblock {LHCb-TDR-015}\relax
\mciteBstWouldAddEndPuncttrue
\mciteSetBstMidEndSepPunct{\mcitedefaultmidpunct}
{\mcitedefaultendpunct}{\mcitedefaultseppunct}\relax
\EndOfBibitem
\bibitem{Scandale:2016krl}
W.~Scandale {\em et~al.},
  \ifthenelse{\boolean{articletitles}}{\emph{{Observation of channeling for
  6500 GeV/c protons in the crystal assisted collimation setup for LHC}},
  }{}\href{http://dx.doi.org/10.1016/j.physletb.2016.05.004}{Phys.\ Lett.\
  \textbf{B758} (2016) 129}\relax
\mciteBstWouldAddEndPuncttrue
\mciteSetBstMidEndSepPunct{\mcitedefaultmidpunct}
{\mcitedefaultendpunct}{\mcitedefaultseppunct}\relax
\EndOfBibitem
\bibitem{Lansberg:2012wj}
J.~P. Lansberg {\em et~al.}, \ifthenelse{\boolean{articletitles}}{\emph{{A
  Fixed-Target ExpeRiment at the LHC (AFTER@LHC) : luminosities, target
  polarisation and a selection of physics studies}}, }{}PoS \textbf{QNP2012}
  (2012) 049, \href{http://arxiv.org/abs/1207.3507}{{\normalfont\ttfamily
  arXiv:1207.3507}}\relax
\mciteBstWouldAddEndPuncttrue
\mciteSetBstMidEndSepPunct{\mcitedefaultmidpunct}
{\mcitedefaultendpunct}{\mcitedefaultseppunct}\relax
\EndOfBibitem
\bibitem{Adare:2006hc}
PHENIX collaboration, A.~Adare {\em et~al.},
  \ifthenelse{\boolean{articletitles}}{\emph{{Measurement of High-${p}_{T}$
  Single Electrons from Heavy-Flavor Decays in $p+p$ Collisions at
  $\sqrt{s}=200\text{ }\mathrm{GeV}$}},
  }{}\href{http://dx.doi.org/10.1103/PhysRevLett.97.252002}{Phys.\ Rev.\ Lett.\
   \textbf{97} (2006) 252002},
  \href{http://arxiv.org/abs/hep-ex/0609010}{{\normalfont\ttfamily
  arXiv:hep-ex/0609010}}\relax
\mciteBstWouldAddEndPuncttrue
\mciteSetBstMidEndSepPunct{\mcitedefaultmidpunct}
{\mcitedefaultendpunct}{\mcitedefaultseppunct}\relax
\EndOfBibitem
\bibitem{Kniehl:2005de}
B.~A. Kniehl and G.~Kramer,
  \ifthenelse{\boolean{articletitles}}{\emph{{${D}^{0}$, ${D}^{+}$,
  ${D}_{s}^{+}$, and ${\ensuremath{\Lambda}}_{c}^{+}$ fragmentation functions
  from CERN LEP1}},
  }{}\href{http://dx.doi.org/10.1103/PhysRevD.71.094013}{Phys.\ Rev.\
  \textbf{D71} (2005) 094013},
  \href{http://arxiv.org/abs/hep-ph/0504058}{{\normalfont\ttfamily
  arXiv:hep-ph/0504058}}\relax
\mciteBstWouldAddEndPuncttrue
\mciteSetBstMidEndSepPunct{\mcitedefaultmidpunct}
{\mcitedefaultendpunct}{\mcitedefaultseppunct}\relax
\EndOfBibitem
\bibitem{Biryukov1997}
V.~M. Biryukov {\em et~al.}, {\em {Crystal Channeling and Its Application at
  High-Energy Accelerators}}, Springer-Verlag Berlin Heidelberg, 1997\relax
\mciteBstWouldAddEndPuncttrue
\mciteSetBstMidEndSepPunct{\mcitedefaultmidpunct}
{\mcitedefaultendpunct}{\mcitedefaultseppunct}\relax
\EndOfBibitem
\bibitem{Jackson:1998nia}
J.~D. Jackson, {\em {Classical Electrodynamics}}, Wiley, 1998\relax
\mciteBstWouldAddEndPuncttrue
\mciteSetBstMidEndSepPunct{\mcitedefaultmidpunct}
{\mcitedefaultendpunct}{\mcitedefaultseppunct}\relax
\EndOfBibitem
\bibitem{Leader2011}
E.~Leader, {\em {Spin in particle physics}}, vol.~15, Camb. Monogr. Part. Phys.
  Nucl. Phys. Cosmol., 2011\relax
\mciteBstWouldAddEndPuncttrue
\mciteSetBstMidEndSepPunct{\mcitedefaultmidpunct}
{\mcitedefaultendpunct}{\mcitedefaultseppunct}\relax
\EndOfBibitem
\bibitem{Fukuyama:2013ioa}
T.~Fukuyama and A.~J. Silenko,
  \ifthenelse{\boolean{articletitles}}{\emph{{Derivation of Generalized
  Thomas-Bargmann-Michel-Telegdi Equation for a Particle with Electric Dipole
  Moment}}, }{}\href{http://dx.doi.org/10.1142/S0217751X13501479}{Int.\ J.\
  Mod.\ Phys.\  \textbf{A28} (2013) 1350147},
  \href{http://arxiv.org/abs/1308.1580}{{\normalfont\ttfamily
  arXiv:1308.1580}}\relax
\mciteBstWouldAddEndPuncttrue
\mciteSetBstMidEndSepPunct{\mcitedefaultmidpunct}
{\mcitedefaultendpunct}{\mcitedefaultseppunct}\relax
\EndOfBibitem
\bibitem{Silenko:2014uca}
A.~J. Silenko, \ifthenelse{\boolean{articletitles}}{\emph{{Spin precession of a
  particle with an electric dipole moment: contributions from classical
  electrodynamics and from the Thomas effect}},
  }{}\href{http://dx.doi.org/10.1088/0031-8949/90/6/065303}{Phys.\ Scripta
  \textbf{90} (2015), no.~6 065303},
  \href{http://arxiv.org/abs/1410.6906}{{\normalfont\ttfamily
  arXiv:1410.6906}}\relax
\mciteBstWouldAddEndPuncttrue
\mciteSetBstMidEndSepPunct{\mcitedefaultmidpunct}
{\mcitedefaultendpunct}{\mcitedefaultseppunct}\relax
\EndOfBibitem
\bibitem{Good:1962zza}
R.~H. Good, \ifthenelse{\boolean{articletitles}}{\emph{{Classical Equations of
  Motion for a Polarized Particle in an Electromagnetic Field}},
  }{}\href{http://dx.doi.org/10.1103/PhysRev.125.2112}{Phys.\ Rev.\
  \textbf{125} (1962) 2112}\relax
\mciteBstWouldAddEndPuncttrue
\mciteSetBstMidEndSepPunct{\mcitedefaultmidpunct}
{\mcitedefaultendpunct}{\mcitedefaultseppunct}\relax
\EndOfBibitem
\bibitem{Metodiev:2015gda}
E.~M. Metodiev, \ifthenelse{\boolean{articletitles}}{\emph{{Thomas-BMT equation
  generalized to electric dipole moments and field gradients}},
  }{}\href{http://arxiv.org/abs/1507.04440}{{\normalfont\ttfamily
  arXiv:1507.04440}}, 2015\relax
\mciteBstWouldAddEndPuncttrue
\mciteSetBstMidEndSepPunct{\mcitedefaultmidpunct}
{\mcitedefaultendpunct}{\mcitedefaultseppunct}\relax
\EndOfBibitem
\bibitem{Grosnick:1989qv}
FNAL-E581/704 collaboration, D.~P. Grosnick {\em et~al.},
  \ifthenelse{\boolean{articletitles}}{\emph{{The design and performance of the
  FNAL high-energy polarized-beam facility}},
  }{}\href{http://dx.doi.org/10.1016/0168-9002(90)90541-D}{Nucl.\ Instrum.\
  Meth.\  \textbf{A290} (1990) 269}\relax
\mciteBstWouldAddEndPuncttrue
\mciteSetBstMidEndSepPunct{\mcitedefaultmidpunct}
{\mcitedefaultendpunct}{\mcitedefaultseppunct}\relax
\EndOfBibitem
\end{mcitethebibliography}

\newpage                                                                                                      


\newpage


\end{document}